\documentclass[journal,comsoc]{IEEEtran}

\usepackage[T1]{fontenc}
\usepackage{cite}
\usepackage{hyperref}

\usepackage{amsmath}
\usepackage{mathtools}
\usepackage[cmintegrals]{newtxmath}

\usepackage{graphicx}
\usepackage{subcaption}
\usepackage{algorithm}
\usepackage{algpseudocode}
\usepackage{multirow}
\usepackage{tablefootnote}

\begin{document}

\title{Planning 5G Networks for Rural Fixed Wireless Access}

\author{Andrew~Lappalainen,
        Yuhao~Zhang,
        and~Catherine~Rosenberg,~\IEEEmembership{Fellow,~IEEE}
\thanks{The authors are with the Department
of Electrical and Computer Engineering, University of Waterloo, Waterloo, ON, Canada}}

\maketitle
\IEEEpeerreviewmaketitle
\begin{abstract}
 We study the planning of a rural 5G multi-user massive MIMO fixed wireless access system to offer fixed broadband service to homes. Specifically, we aim to determine the user limit, i.e., the maximum number of homes that can simultaneously receive target minimum bit rates (MBRs) on the downlink (DL) and on the uplink (UL) given a set of network resources and a cell radius. To compute that limit, we must understand how resources should be shared between the DL and UL and how user and stream selection, precoding and combining, and power distribution should be performed. We use block diagonalization and propose a static grouping strategy that organizes homes into fixed groups (of possibly different sizes) in the DL and UL; then we develop a simple approach to compute the user limit that we validate numerically. We study the impact of group size and show that smaller groups yield larger user limits in a 3.5~GHz band. We show how the user limit at different cell radii is impacted by the system bandwidth, the number of antennas at the base station and homes, the transmit power, and the DL and UL MBRs. Lastly, we offer insights into how the network could be operated.
\end{abstract}


\section{Introduction}

Rural Internet users continue to suffer from poor broadband due to the high costs to build wired network infrastructure relative to the low density of users, resulting in a digital divide between rural and urban communities. Historically, mobile and home access have been provided through different technologies, with mobile services coming from cellular technologies and fixed (home) broadband coming from fixed access technologies. However, a convergence between mobile and home access has started to emerge through cellular fixed wireless access (FWA) technology, making a cost-effective alternative for fixed broadband (FB). Initial 4G-based FWA deployments were unable to compete with wired FB performance, but the advancements 5G brings via multi-user massive MIMO (MU-MIMO)~\cite{larsson2014mimo}, a flexible new radio (NR)~\cite{3gppTS38211}, and new spectrum~\cite{shafi2017tutorial} make 5G FWA a promising alternative for FB, especially in rural areas~\cite{lappalainen2022comstd}.

FWA FB services differ from the types of services normally offered over mobile networks. In particular, FB has much higher data rate requirements, not only on the downlink (DL), which is necessary for streaming high-definition video content on large screens, but on the uplink (UL) as well, to enable users to participate in virtual meetings or small businesses to host servers. Hence, a key challenge mobile network operators (MNOs) face is to ensure that every subscribing home receives a \emph{minimum bit rate (MBR)} in the DL \emph{and} in the UL (although they might be different). MBRs are typically offered with other FB technologies, but in wireless environments this requirement is much more difficult to achieve since the channel changes with the user location and the environment. Nonetheless, since homes are fixed (non-mobile) and could be equipped with large multi-antenna arrays that enable simultaneous transmission of multiple streams, MNOs could use 5G FWA to help bridge part of the digital divide in rural areas.

MNOs offering a rural 5G FWA service with DL and UL MBRs must ensure that their networks are properly planned and dimensioned. Given the high 4G coverage that already exists in rural communities, it is anticipated that early 5G FWA deployments can make use of existing towers, thus avoiding the base station (BS) placement problem. Thus, the initial planning step MNOs must take is to determine the number of homes that can be offered the DL and UL MBRs in this context. 

Specifically, in this paper, we consider a single cell 5G MU-MIMO system using a 3.5~GHz\footnote{Though our methodology is agnostic to the  band used.} time-division duplex (TDD)\footnote{TDD allocates the full bandwidth to DL transmission for a fraction of time and to UL transmission for the remaining fraction of time.} band, in which homes have multiple antennas supporting multi-stream transmission and reception. Although 5G FWA has started to be deployed in some urban markets using mmWave spectrum, it is unclear how effectively mmWave can be operated over large rural areas. On the other hand, new mid-band 5G spectrum, such as 3.5~GHz, is expected to be widely adopted for rural FWA~\cite{schumacher2019coverage}. Inter-cell interference is expected to be low in a rural environment, so we focus on a single cell scenario. Our first objective is to understand the user limit (i.e., the maximum number of rural FWA homes that can be offered the DL and UL MBRs simultaneously) for a given system setting, i.e., a cell radius, bandwidth, DL and UL MBRs, number of antennas at the BS and homes, and power at the BS and homes. Because the system is TDD, DL and UL resources are shared in time, where the UL gets a certain number of time-slots in a frame, and the DL gets the remaining slots. For this reason, the DL and UL problems cannot be studied independently: this is a joint problem in which it is necessary to determine how many resources are used for the DL vs the UL. Hence, a related objective is to determine the right number of time-slots to allocate to the UL and DL to attain the user limit.

To determine the user limit, we need to mimic the operation of the network (since it impacts performance), i.e., take different radio resource management  procedures into account, such as user selection (which homes and streams get scheduled), precoding and combining (the beamforming at the transmitters and receivers, respectively), the power distribution (PD) (distribution of power to each stream), and  the modulation and coding (MCS) selected per stream (which determines the data rate based on the channel conditions). Specifically, to answer the question: ``can $U$ homes be supported with a high probability?", we would need to generate many different realizations  with $U$ homes (i.e., random distributions of their locations) and check, for each realization, if there is a way to offer each home the DL and UL MBRs. To do that for a given randomly distributed set of homes and their corresponding channel matrices, a problem that jointly optimizes user selection, precoding and power distribution, can be formulated to determine if there is a way to offer the DL and UL MBRs to all of the homes. However, the resulting problem is very large and complex, and to obtain meaningful user limits, this problem would need to be solved for many different numbers of homes $U$ for many different realizations.

To simplify the problem we adopt block diagonalization (BD) precoding and combining in both the UL and DL. BD limits the number of homes scheduled on the same Physical Resource Block (PRB) to $N = \left \lceil \frac{M_{\rm BS}}{M_{\rm H}} \right \rceil$~\cite{shen2006low} (where $M_{\rm BS}$ is the number of antennas at the BS and $M_{\rm H}$ is the number of antennas at each home). $N$ could be small in an FWA system where homes have multiple antennas. The expectation is that MNOs will be able to support a number of homes much larger than $N$ and hence, we \emph{cannot} assume that all homes are scheduled in each PRB and we must adopt a different strategy to properly select users. We adopt a static user grouping strategy, which organizes homes into fixed groups no larger than $N$ and allocates an equal share of PRBs to each group (the group sizes could be different on the DL and UL). The static grouping strategy enables us to compute a user limit in a simple and robust manner. It is possible that in an operational phase a more flexible user selection strategy would be more efficient.

To determine whether a given realization of $U$ users can be given the DL (resp. UL) MBR, we solve a convex nonlinear programming problem (NLP) for the DL (resp. UL) PD problem, where the objective is to maximize the \emph{minimum} data rates given to the users during a frame given a power budget, a grouping of users under BD, the channel conditions, and a known MCS function, as well as a certain (TDD) partitioning between the DL and UL. If the minimum date rate meets or exceeds the DL (resp. UL) MBR for 95\% of realizations then the DL (resp. UL) MBRs can be given to U users. We say that $U$ users can be given the MBRs if there exists a time partition where the DL \emph{and} UL MBRs can be given to $U$ users. To compute the user limit we could keep solving these problems for successively larger numbers of users $U$; however (as stated in Contribution 2 below), we determined a more efficient way to compute the user limit using our convex NLP problem.

We now summarize our contributions:

\emph{Contribution 1:} In the context of computing user limits through grouping, we first need to understand if we should always select $N$ homes per PRB and how many streams should be selected per home. Our first contribution is to show that selecting $N$ (or close to $N$) homes per PRB is generally not optimal. In fact, a greater number of homes can be given the MBRs (i.e., a larger user limit can be attained) if we select fewer homes with more streams than more homes with fewer streams.

\emph{Contribution 2:} By adopting the static grouping strategy, we develop a simple approach to determine the user limit for a given system setting. First, we (separately) determine the minimum number of subchannels needed to provide the DL and UL MBRs for different group sizes for a given (TDD) partition of the time between the UL and DL. Then, knowing the total number of subchannels available in the system, we use this information to determine how many groups can be given the MBRs on the UL and on the DL, and, hence, how many homes can be given both MBRs for the given TDD partition. We repeat this exercise for different TDD partitions and group sizes (a relatively small range of combinations) and determine the configuration that maximizes the number of homes that can be supported. The elegance of this approach is that it enables MNOs to quickly plan their network for a MU-MIMO FWA system with MBRs requirements. 

\emph{Contribution 3:} We validate our approach against brute force computation. Then, we use our approach to show how the user limit at different cell radii is impacted by the following: the system bandwidth, the number of antennas at the BS and at the homes, the BS transmit power and the DL and UL MBRs.

\emph{Contribution 4:} Although our main focus is network planning, the results obtained can also indicate to MNOs which TDD partition and DL/UL group sizes are optimal when the number of active homes is less than the user limit. In other words, these results can provide insight into how the network could be operated. Specifically, we show that by adjusting the group sizes and the TDD partition according to the number of active homes, the DL and UL rates given to homes can be improved compared to when the network is operated statically using the configuration that is required for the user limit.

The tools that we have developed and the results that we have obtained can help MNOs decide if introducing FWA for fixed broadband is profitable, i.e., if the user limits are large enough to warrant the implementation of that service as well as help them provision their network, i.e., decide the bandwidth, power, number of antennas needed for a target user limit.

We structure the remainder of this paper as follows. In Section~\ref{sect:related_work} we review the related work. We describe the system model in Section~\ref{sect:system_model}. The problem formulation and the approach to determine the user limits are presented in Section~\ref{sect:problem_formulation}. In Section~\ref{sect:results} we provide the numerical results, which validate our proposed approach, answer the network planning questions and provide preliminary insights to some network operation questions. We conclude the paper in Section~\ref{sect:conclusion}.

\textit{Notation}: $\mathbb{R}$ and $\mathbb{C}$ denote the set of real and complex numbers, respectively. $\mathbb{R}_+$ denotes the set of non-negative real numbers. Vectors and matrices are shown by bold lower-case and bold upper-case letters, respectively.

\section{Related Work}\label{sect:related_work}

Recent work on multi-user massive MIMO for FWA was presented in~\cite{colpaert2020fixed,brighente2020modular}; however, the focus was on beamforming. Moreover, these papers assumed that the number of users were $U \leq N$ and considered mmWave bands only. In fact, many papers on 5G FWA assume the use of mmWave bands, even in rural areas~\cite{aldubaikhy2020mmwave,abozariba2019evaluating,kaddoura2019greenfield}. However, mmWave limits the cell radius (e.g.~\cite{abozariba2019evaluating} states that mmWave cannot practically provide FWA coverage beyond 1500~m under BS transmission power regulations), which is why we focus on mid-bands.

The authors of \cite{jaber2015tutorial} and \cite{taufique2017planning} provide an overview of network dimensioning and planning in 4G while looking ahead to 5G. They also summarize a number of classical references on these topics. More recently, \cite{kaddoura2019greenfield}~considered the BS placement problem to guarantee 5G FWA homes a minimum coverage level and DL data rate using mmWave cells. \cite{chiaraviglio2021planning}~also modeled the 5G BS placement problem to guarantee a minimum DL service level while simultaneously minimizing BS electromagnetic field power. In our problem, we assume that MNOs will reuse the existing 4G towers for 5G FWA, hence our focus is on quantifying the number of FWA homes that can be provided a MBR in both the DL and UL. Another recent paper used crowdsourced data to dimension existing BSs and determine the appropriate per-site NR configurations required to support new 5G services~\cite{khan2020service}; however, their dimensioning process was based on peak DL data rate requirements.

Although none of these recent works considered the UL in their problems, some earlier work on 4G planning and dimensioning considered DL and UL service requirements jointly. \cite{kashef2016balanced}~examined how 4G BSs could be planned to provide DL and UL MBRs while supporting ON-OFF switching to save energy. However, user limits were assumed a priori. Furthermore, their system was FDD and had fixed DL/UL bandwidths, and thus, did not consider the resource allocation problem. \cite{ghazzai2015optimized}~studied the dimensioning and planning problems for 4G networks with DL and UL MBRs, but the user limits computed in their dimensioning stage only relied on DL MBRs and the spectral efficiency was assumed to be constant throughout the cell. Furthermore, neither of these papers considered MU-MIMO.

In the past, the problem of providing MBRs in MU-MIMO systems has been studied separately for the DL \cite{ng2012energy,lima2013improved} and the UL \cite{zappone2015distributed,zappone2015energy}. The authors of~\cite{ho2008optimal} formulated the problem for both the DL and the UL of a MU-MIMO system but considered the two directions in isolation, thus ignoring the issue of providing MBRs to the DL and UL simultaneously.

Outside of the MIMO context, some earlier work studied how to jointly provide DL and UL QoS guarantees in TDD-based systems. \cite{elhajj2011dynamic}~examined the problem of dynamic resource sharing in TDD OFDMA networks with symmetric DL and UL MBR requirements; and \cite{zorba2015energy}~studied the problem of providing minimum DL and UL SINR levels to users in a multi-cell TDD-LTE system where cells may have different time-slot configurations.

To the best of our knowledge, no previous work has examined the problem of planning a network while optimizing the resource configuration to jointly provide a minimum data rate for the DL and UL in an OFDM-based MU-MIMO system. Through the application of user grouping, our problem is generalized to the case where the total number of homes $U$ is such that $U > N$ where $N = \left \lceil \frac{M_{\rm BS}}{M_{\rm H}} \right \rceil$ is the ratio of $M_{\rm BS}$ the number of antennas at the BS to $M_{\rm H}$ the number of antennas at each home (though it could just as easily be applied when $U \leq N$). Although many papers modeling MU-MIMO systems assume that the total number of users $U$ is such that $U \leq N$ and, hence, all users can be selected simultaneously~\cite{colpaert2020fixed,brighente2020modular,ng2012energy,zappone2015distributed,zappone2015energy} (also, \cite{castaneda2016overview} and references therein), earlier work on MU-MIMO systems where the total number of antennas at the BS was assumed to be smaller recognized the need for user selection under zero-forcing (ZF)~\cite{yoo2006optimality} and BD~\cite{shen2006low} precoding but they only considered the downlink, and did not impose minimum rate requirements for users nor explore the impact different group sizes have on achieving minimum rates.

\section{System Model and Radio Resource Management Processes}\label{sect:system_model}
We consider a DL and UL MU-MIMO system, with a BS having $M_{\rm BS}$ antennas and a certain number of homes each equipped with a 5G modem and $M_{\rm H}$ rooftop antennas. The system is TDD. A frame is composed of $T$ slots in the time-domain, $T_d$ allocated to the DL and $T_u$ allocated to the UL ($T_d + T_u = T$). In the frequency domain the system has a bandwidth $B$ made up of a total of $C$ subchannels\footnote{We use the terms channel and subchannel interchangeably in the paper.}, each with bandwidth $B_C$. A PRB is composed of one subchannel and one time-slot, and, hence, there are $C \times T$ PRBs in a frame. The BS transmits with power $P_{\rm max}$ on the DL and each FWA home transmits with power $P_{\rm H}$ on the UL. On the DL, the power budget is shared equally by the PRBs in a given time-slot, i.e., $P_{PRB}^d=\frac{P_{\rm max}}{C}$. On the UL, the power allocated to a PRB by home $k$ is $P_{PRB}^k=\frac{P_{\rm H}}{C_k}$ where $C_k \leq C$ is the number of channels allocated to home $k$ (to be detailed later). Since we consider a rural environment with little to no inter-cell interference, we adopt a single-cell circular topology with a radius $\mathcal{R}$. In the following we refer to $$\mathcal{S} = (B,C,P_{\rm max},P_{\rm H},MBR_d,MBR_u, M_{\rm BS},M_{\rm H})$$ as the system \textit{setting}. Next, we describe the radio resource management processes.

\subsection{User Selection}
For a given setting $\mathcal{S}$, a cell radius $\mathcal{R}$, a given realization $\omega(U)$ (which is characterized by a randomly distributed set of $U$ FWA homes and their channel matrices as discussed later), user selection is performed once for all time, separately in the DL and the UL. We adopt a grouping strategy that divides homes into equal-sized groups of $S \leq N$ homes (let $S_d$ (resp. $S_u$) be the group size on the DL (resp. UL)). Because grouping is done separately for the DL and UL, $S_d$ and $S_u$ might be different. In this paper, we adopt a random strategy when grouping homes. Smarter grouping strategies are for further study.

In the following we refer to the triple $\mathcal{V}=(T_u,S_d,S_u)$ as the system \textit{configuration} and the pair $\mathcal{V}_d=(T_u,S_d)$ (resp. $\mathcal{V}_u=(T_u,S_u)$) as the DL (resp. UL) system configuration. The result of the user selection process for a configuration~$\mathcal{V}$ and a realization $\omega(U)$, is the creation of static groups of homes.

For a given system setting $\mathcal{S}$ and a radius $\mathcal{R}$, we consider a large set $\Omega(U)$ of realizations with $U$ homes. If more than 95\% of the realizations can offer (at least) $MBR_d$ and $MBR_u$ to their $U$ homes with configuration $\mathcal{V}$, we say that the configuration $\mathcal{V}$ is $U$-feasible. Note that there might be many feasible configurations for $U$ for a setting $\mathcal{S}$ and a radius $\mathcal{R}$.

\subsection{Channel Allocation}
From above, the user selection process divides the homes in the realization $\omega(U)$ into static groups of size $S_d$ in the DL and $S_u$ in the UL. In the DL there is a total of $G(S_d) = \left\lceil \frac{U}{S_d} \right\rceil$ groups $\hat{\mathcal{U}}_1,\ldots,\hat{\mathcal{U}}_{G(S_d)}$, and each group $\hat{\mathcal{U}}$ is randomly assigned a set $\mathcal{C}_d(\hat{\mathcal{U}})$ of $n(S_d) = \left\lfloor \frac{C}{G(S_d)} \right\rfloor$ subchannels (note that there might be $n' < n(S_d)$ leftover subchannels, in which case one extra subchannel is given to the first $n'$ groups). For the UL there is a total of $G(S_u) = \left\lceil \frac{U}{S_u} \right\rceil$ groups $\hat{\mathcal{U}}_1,\ldots,\hat{\mathcal{U}}_{G(S_u)}$, each of which is randomly assigned a set $\mathcal{C}_u(\hat{\mathcal{U}})$ of $n(S_u) = \left\lfloor \frac{C}{G(S_u)} \right\rfloor$ subchannels. Note that both user selection and channel allocation are static, i.e., they do not change from one frame to another.

\subsection{Precoding and Combining Schemes}

Given a setting $\mathcal{S}$, a radius $\mathcal{R}$, a configuration $\mathcal{V}$, and a realization $\omega(U)$, we perform precoding and combining independently for the UL and DL once at the start of each frame for each distinct group of homes created during user selection, for each of the channels allocated to the groups (because the channels are not assumed flat in frequency, see next paragraph). Note that the maximum number of streams at each home is $M_{\rm H}$. At this stage, we assume that there are $L=M_{\rm H}$ streams per home. Note that PD might not allocate enough power to some streams to have a non-zero rate in some PRBs (see Section~\ref{sect:problem_formulation}) and hence the number of active streams might be less than $M_{\rm H}$.

Full channel state information (CSI) is assumed. We represent the channel matrix between the BS and the \mbox{$k$-th} home at PRB~$(c,t)$ as $\mathbf{G}_k^{c,t} = \mathbf{G}_k^c$. Since each home is fixed, the channel coherence time is large; hence, we assume that the channel coefficients are constant throughout a frame. On the other hand, we do not assume the channel is flat in frequency (i.e., it is frequency-selective). Furthermore, we assume that the channel is the same in the DL and UL because the system is TDD. Note that the channel model itself does not impact the system model and problem formulation. Hence, we leave the details of the channel model to Section~\ref{sect:channel}.

We adopt a normalized BD precoding and combining process, which enables us to decouple precoding and combining (a computation that does not require any optimization) from power distribution. Furthermore, since we assume equal power allocation for each subchannel, we may perform precoding and combining independently per PRB. Because subchannels are time-invariant within a frame and a subchannel is allocated to the same group of homes within a frame, we only need to perform precoding/combining once per subchannel per frame. Hence, given a group of $\hat{U} \leq N$ homes, denoted by $\hat{\mathcal{U}}$, is selected for a subchannel $c$ in a frame, under BD, the normalized precoding matrix at the BS for home $k$ in subchannel $c$ given by $\mathbf{W}_k^c = \left[ \mathbf{w}_{k,1}^c, \ldots, \mathbf{w}_{k,L}^c \right] \in \mathbb{C}^{M_{\rm BS} \times L}$ is used to transmit data symbols to each stream $l \in \{1,2,\ldots,L\}$, and the combining matrix at each home given by $\mathbf{U}_k^c = \left[ \mathbf{u}_{k,1}^c, \ldots, \mathbf{u}_{k,L}^c \right]^T \in \mathbb{C}^{L \times M_{\rm H}}$ is used to recover the data symbols. More details about BD can be found in \cite{shen2006low,spencer2004zero,chae2008block,zu2013generalized}.

The output of the normalized precoding and combining process is a set of effective channels $\{E_{k,l}^c\}$ for each (possible) user stream $l \in \{1,\ldots,L\}$ of home $k \in \hat{\mathcal{U}}$ for each subchannel $c$.

Next, we outline the steps needed to compute the effective channels for a single subchannel on the DL for a given group of homes $\hat{\mathcal{U}}$. For brevity, we omit the per-subchannel superscript $c$ in the following. Due to symmetry in the channel, the UL procedure is the same as the DL procedure, except that the precoding matrix at the BS in the DL becomes the combining matrix at the BS in the UL and the combining matrix at the home in the DL becomes the precoding matrix at the home in the UL. The resulting effective channels are the same for the UL and DL.

Let $\mathbf{s}_k = [s_{k,1}, s_{k,2}, \ldots, s_{k,L}] \in \mathbb{C}^{L \times 1}$ be the dedicated data symbol vector for the \mbox{$k$-th} home, where $s_{k,l}$ is the data symbol for stream $l$, $\mathbb{E}\{ \mathbf{s}_k \mathbf{s}_k^H \} = \mathbf{I}_{L}$, $k \in \hat{\mathcal{U}}$ and $\mathbf{I}_{L}$ is the $L \times L$ identity matrix. The recovered data symbols at home $k$ are given by
\begin{equation}\label{eq:RecoveredSymbols}
\hat{\mathbf{s}}_k = \mathbf{U}_k \mathbf{G}_k \mathbf{W}_k \mathbf{P}_k \mathbf{s}_k + \underbrace{\sum_{i \in \hat{\mathcal{U}}, i \neq k} \mathbf{U}_k \mathbf{G}_k \mathbf{W}_i \mathbf{P}_i \mathbf{s}_i}_{\rm inter-user\, interference} + \mathbf{U}_k \mathbf{z}_k, \quad k \in \hat{\mathcal{U}},
\end{equation}
where $\mathbf{G}_k = \left[ \mathbf{g}_{k}^1, \ldots, \mathbf{g}_{k}^{M_{\rm H}} \right]^T \in \mathbb{C}^{M_{\rm H} \times M_{\rm BS}}$ is the channel matrix between the BS and home, $\mathbf{P}_k \in \mathbb{R}_+^{L \times L}$ is a diagonal matrix representing the power distribution for the data streams of the \mbox{$k$-th} home, and $\mathbf{z}_k \sim \mathcal{CN} (0,\sigma^2 \cdot \mathbf{I}_{M_{\rm H}})$ is the additive white Gaussian noise~(AWGN). 

The key idea behind BD precoding is to select the precoding matrix to fully suppress the inter-user interference, i.e.,
\begin{equation}
    \mathbf{G}_k \mathbf{W}_i \mathbf{P}_i = \textbf{0}, \quad \forall i,k \in \hat{\mathcal{U}}, i \neq k.
\end{equation}
To do so, let $\widetilde{\mathbf{G}}_k$ represent the channel matrix from the BS to all homes except the \mbox{$k$-th} home, i.e., $\widetilde{\mathbf{G}}_k = [\mathbf{G}_1^T,\ldots,\mathbf{G}_{k-1}^T,\mathbf{G}_{k+1}^T, \ldots,\mathbf{G}_{\hat{U}}^T]^T$. By conducting singular value decomposition~(SVD) for $\widetilde{\mathbf{G}}_k$, we have
\begin{equation}
    \widetilde{\mathbf{G}}_k = \widetilde{\mathbf{T}}_k \widetilde{\mathbf{\Lambda}}_k [\widetilde{\mathbf{V}}_k^1 \quad \widetilde{\mathbf{V}}_k^0]^H, \quad k \in \hat{\mathcal{U}},
\end{equation}
where $\widetilde{\mathbf{T}}_k \in \mathbb{C}^{(\hat{U} - 1)M_{\rm H} \times (\hat{U} - 1)M_{\rm H}}$ is a unitary matrix; $\widetilde{\mathbf{\Lambda}}_k \in \mathbb{R}_+^{(\hat{U} - 1)M_{\rm H} \times M_{\rm BS}}$ a rectangular diagonal matrix; $\widetilde{\mathbf{V}}_k^1 \in \mathbb{C}^{M_{\rm BS} \times (M_{\rm BS} - R_k)}$ and $\widetilde{\mathbf{V}}_k^0 \in \mathbb{C}^{M_{\rm BS} \times R_k}$ are the submatrices composed of the right-singular vectors corresponding to non-zero and zero singular values, respectively, where $R_k > 0$ is the dimension of the nullspace of $\widetilde{\mathbf{G}}_k$. (In order to have zero inter-user interference, $\mathbf{W}_k \mathbf{P}_k$ must lie in the nullspace of $\widetilde{\mathbf{G}}_k$; hence, the null space of $\widetilde{\mathbf{G}}_k$ must have a dimension that is greater than 0.) From this, we obtain the non-interfering block effective channel matrix
\begin{equation}\label{eq:BlockEffChannelMatrix}
    \mathbf{G}_k^e = \mathbf{G}_k \cdot \widetilde{\mathbf{V}}_k^0, \quad k \in \hat{\mathcal{U}}
\end{equation}
Through \eqref{eq:BlockEffChannelMatrix}, the MU-MIMO system reduces into $\hat{U}$ parallel and non-interfering single-user MIMO (SU-MIMO) systems. From this, we can separately derive the (normalized) precoding and combining per user stream $(k,l)$. To do so, we perform SVD on the effective channel matrix \eqref{eq:BlockEffChannelMatrix} for every home $k$~\cite{zu2013generalized}
\begin{equation}
    \mathbf{G}_k^e = \mathbf{T}_k^e \cdot \mathbf{\Lambda}_k^e \cdot (\mathbf{V}_k^e)^H, \quad k \in \hat{\mathcal{U}},
\end{equation}
where $\mathbf{T}_k^e \in \mathbb{C}^{M_{\rm H} \times M_{\rm H}}$ and $\mathbf{V}_k^e \in \mathbb{C}^{R_k \times R_k}$ are unitary matrices, and $\mathbf{\Lambda}_k^e \in \mathbb{R}_+^{M_{\rm H} \times R_k}$ is a rectangular diagonal matrix containing the singular values of $\mathbf{G}_k^e$ sorted in descending order

\begin{equation}
    \mathbf{\Lambda}_k^e = 
    \begin{bmatrix}
    \lambda_{k,1} & \cdots & & \mathbf{0} \\
    & \ddots & & \\
    \mathbf{0} & \cdots & \lambda_{k,L} & \mathbf{0} 
    \end{bmatrix}
\end{equation}

where $\lambda_{k,l} = \left| \mathbf{u}_{k,l} \mathbf{G}_k \mathbf{w}_{k,l} \right|$ is the normalized precoding and combining on stream $(k,l)$.
Thus, the effective channel for stream $l \in \{1,\ldots,L\}$ of home $k \in \hat{\mathcal{U}}$ is
\begin{equation}
    E_{k,l} = \frac{\lambda_{k,l}^2}{\sigma^2}.
\end{equation}
Recall that there is an effective channel per stream per home in the group, per subchannel allocated to that group.

\subsection{Power Distribution}
Given a setting $\mathcal{S}$, a radius $\mathcal{R}$, a configuration $\mathcal{V}$, a direction (either UL or DL), a realization $\omega(U)$, a grouping of homes obtained via user selection and their channel, and the per stream effective channels for each home in the groups, our next challenge is to verify if the configuration is feasible, i.e., if there exists a per stream PD that enables each selected home to receive its MBRs in the DL and UL directions. At this stage a stream that does not receive enough power to see a non-zero rate in a PRB can be considered as not being selected. We cast the PD problem independently for each group as a max-min problem over all the PRBs allocated to the group as presented in Section~\ref{sect:problem_formulation}.

The outcome of PD is the power per PRB per stream per home. With these powers, we can compute the SINR for each stream of each home of a group in each PRB belonging to that group.

\subsection{Rate Computation Using Modulation and Coding Scheme}
A modulation and coding scheme defines the spectral-efficiency function which maps the SINR to a spectral efficiency. In contrast to most of the earlier papers providing MBRs that utilize the Shannon capacity~\cite{chiaraviglio2021planning,kashef2016balanced,ghazzai2015optimized,ng2012energy,zappone2015distributed,zappone2015energy,ho2008optimal,elhajj2011dynamic}, we use a practical MCS function which is piecewise constant and characterized by $Q$ pairs $(\text{SE}_q,\gamma_q)$, i.e., if $\gamma_{q} \leq SINR_{k,l}^{c,t} < \gamma_{q+1}$, then the spectral efficiency seen by the \mbox{$l$-th} data stream is $\text{SE}_q$, with units bps/Hz (where $\gamma_1<\gamma_2<\cdots$ and $\text{SE}_1<\text{SE}_2<\cdots$). Then, given the spectral efficiency $\text{SE}_q$ and the subchannel bandwidth $B_C$, the outcome of this process is the rate seen by each home in each PRB, i.e., the sum of the rates seen by each stream of that home in that PRB. More details on the MCS function are given in Section~\ref{sect:channel}.

\section{PD Problems Formulation and User Limit Computations}\label{sect:problem_formulation}
In this section, we focus on the power distribution problems as well as the procedure to compute the user limit. These steps are called after user selection (i.e., grouping), channel allocation and precoding and combining are performed. Recall that the UL and DL PD problems can be decoupled by fixing the number of slots $T_u$ allocated to the UL and $T_d = T - T_u$ allocated to the DL. In determining the UL and DL user limits for all possible values of $T_u$, we will determine the optimal $T_u$ (the value that maximizes the minimum of the UL and DL limit).

We work with a full buffer assumption. In other words, we assume that each of the homes is greedy and there is an infinite supply of bits at the cell to be sent to the homes in the DL and an infinite supply of bits at the homes to be sent to the BS in the UL. The full buffer assumption enables us to work with rates without needing to consider short-term traffic fluctuations that could otherwise cause the buffer to be empty or insufficiently full to meet the MBR in a given frame. This model works well in a planning stage for understanding the achievable user limit for a given system setting and a given radius and the configuration required to achieve that limit. We will first formulate the problem for the DL; next, we will formulate the UL problem; and then we will explain how to obtain the user limits.

\subsection{Downlink Formulation}
For a given setting and radius, for a given realization $\omega(U)$, a given DL configuration $\mathcal{V}_d=(T_u,S_d)$, the downlink user selection (i.e., user grouping) process divides the homes into static groups of size $S_d$. There is a total of $G(S_d) = \left\lceil \frac{U}{S_d} \right\rceil$ groups $\hat{\mathcal{U}}_1,\ldots,\hat{\mathcal{U}}_{G(S_d)}$\footnote{Possibly some of the groups could have $S_d-1$ homes instead of $S_d$.}, and each group $\hat{\mathcal{U}}$ is randomly assigned a set $\mathcal{C}_d(\hat{\mathcal{U}})$ of $n(S_d) = \left\lfloor \frac{C}{G(S_d)} \right\rfloor$ subchannels throughout the frame. Given $L\leq M_{\rm H}$ streams per home, the per-stream effective channels are computed for each home in the group and for each subchannel in $\mathcal{C}_d(\hat{\mathcal{U}})$. The reason why we might choose $L<M_{\rm H}$ is because we might be interested in forcing the number of streams to be small to see the impact they have on performance. This is what we do in Section~\ref{toto}.

Given the downlink piecewise constant MCS function with a set $\mathcal{Q}$ of $Q$ levels, a DL configuration $\mathcal{V}_d$, a group $\hat{\mathcal{U}}$ and its allocated subchannels $\mathcal{C}_d(\hat{\mathcal{U}})$, knowing the effective channels $\{E_{k,l}^c\}$, $k \in \hat{\mathcal{U}}$, $l \in \{1,\ldots,L\}$, $c \in \mathcal{C}_d(\hat{\mathcal{U}})$, the objective is to verify whether there exists a distribution of power $\{P_{k,l}^{c,t}\}$ to each user stream in each PRB corresponding to channels in $\mathcal{C}_d(\hat{\mathcal{U}})$ so that we can give $MBR_d$ to the homes in $\hat{\mathcal{U}}$. Let $\mathcal{T}_d$ be the set of DL time-slots, where $|\mathcal{T}_d| = T_d = T - T_u$. We formulate the DL power distribution problem $\mathbf{P_d^0}(\hat{\mathcal{U}},\mathcal{V}_d,L)$ in a given frame as: 

\begin{align}
    & \max_{\lambda(\hat{\mathcal{U}},\mathcal{V}_d,L),(P_{k,l}^{c,t}), (r_{k,l}^{c,t}), (\lambda_k), (I_{k,l,q}^{c,t})} \lambda(\hat{\mathcal{U}},\mathcal{V}_d,L) \label{eq:Objective} \\
    & \lambda_k \geq \lambda(\hat{\mathcal{U}},\mathcal{V}_d,L) \quad \forall k \in \hat{\mathcal{U}} \label{eq:MinRate} \\
    & \lambda_k = \frac{1}{T} \sum_{c \in \mathcal{C}_d(\hat{\mathcal{U}})} \sum_{t \in \mathcal{T}_d}  \sum_{l=1}^{L} r_{k,l}^{c,t} \quad \forall k \in \hat{\mathcal{U}} \label{eq:UserRate}\\
    & \sum_{k \in \hat{\mathcal{U}}} \sum_{l=1}^{L} P_{k,l}^{c,t} \leq P_{PRB}^d = \frac{P_{\rm max}}{C}  \quad \forall c \in \mathcal{C}_d(\hat{\mathcal{U}}), t \in \mathcal{T}_d \label{eq:DLPowerConst} \\
    & \gamma_q I_{k,l,q}^{c,t} \leq P_{k,l}^{c,t} E_{k,l}^c \label{eq:SINRLevel} \quad \forall c,t,k,l \in \{1,\ldots,L\},q \in \mathcal{Q} \\ 
    & r_{k,l}^{c,t} = B_C \sum_{q \in \mathcal{Q}} \text{SE}_q I_{k,l,q}^{c,t} \quad  \forall c,t, k,l \label{eq:StreamRate} \\
    & \sum_{q \in \mathcal{Q}} I_{k,l,q}^{c,t} \leq 1 \quad  \forall c,t, k,l \\
    & I_{k,l,q}^{c,t} \in \{0,1\} \quad \forall c,t, k, l,q \label{eq:IntVar}
\end{align}

Through~\eqref{eq:Objective} and~\eqref{eq:MinRate}, we seek to maximize the minimum user rate in the group, $\lambda(\hat{\mathcal{U}},\mathcal{V}_d,L)$. \eqref{eq:UserRate} defines the rate $\lambda_k$ given to home $k$ in the frame, which is a function of the per-PRB, per-stream rates $r_{k,l}^{c,t}$. \eqref{eq:DLPowerConst} constrains the power to use on a PRB. This problem uses the practical MCS function, where the selected MCS level $q$ is determined by~\eqref{eq:SINRLevel} and the resulting rate $r_{k,l}^{c,t}$ is given in~\eqref{eq:StreamRate}. The practical MCS function is piecewise constant and introduces integer variables $I_{k,l,q}^{c,t}$ (see~\eqref{eq:IntVar}), making the problem a mixed integer nonlinear programming problem (MINLP). This piecewise function has a derivative of 0 everywhere except at the points of discontinuity, which makes the optimization problem $\mathbf{P_d^0}(\hat{\mathcal{U}},\mathcal{V}_d,L)$ difficult to solve~\cite{boyd2004convex}.

To make the problem more tractable, we approximate the MCS function with a continuous function. Specifically, the rate given to the \mbox{$l$-th} stream of home~$k$ in PRB~$(c,t)$ can be approximated by~\cite{zhang2021joint} \begin{equation}\label{eq:ApproxMCS}
    r_{k,l}^{c,t} = B_C \min (SE_{\rm max}, a(SINR_{k,l}^{c,t})^b)
\end{equation}
where $a > 0$, $0 \leq b \leq 1$, $SE_{\rm max}$ is the highest SE of the practical MCS function and $SINR_{k,l}^{c,t} = P_{k,l}^{c,t} E_{k,l}^c$. We denote $r_{max} = B_C SE_{\rm max}$ in the following. Note that this function is continuous and gives a non-zero rate for any $SINR>0$ while the practical MCS function is strictly positive only if $SINR \geq \gamma_1$ where $\gamma_1$ is the first threshold of the MCS function.

With \eqref{eq:ApproxMCS}, $\mathbf{P_d^0}(\hat{\mathcal{U}},\mathcal{V}_d,L)$ can be approximated by $\mathbf{P_d}(\hat{\mathcal{U}},\mathcal{V}_d,L)$:
\begin{align}
    & \max_{\lambda(\hat{\mathcal{U}},\mathcal{V}_d,L),(P_{k,l}^{c,t}), (r_{k,l}^{c,t}), (\lambda_k)} \lambda(\hat{\mathcal{U}},\mathcal{V}_d,L) \\
    & \lambda_k \geq \lambda(\hat{\mathcal{U}},\mathcal{V}_d,L) \quad \forall k \in \hat{\mathcal{U}} \\
    & \lambda_k = \frac{1}{T} \sum_{c \in \mathcal{C}_d(\hat{\mathcal{U}})} \sum_{t \in \mathcal{T}_d}  \sum_{l=1}^{L} r_{k,l}^{c,t} \quad \forall k \in \hat{\mathcal{U}}\\
    & \sum_{k \in \hat{\mathcal{U}}} \sum_{l=1}^{L} P_{k,l}^{c,t} \leq P_{PRB}^d = \frac{P_{\rm max}}{C} \quad \forall c \in \mathcal{C}_d(\hat{\mathcal{U}}), t \in \mathcal{T}_d
     \label{eq:dl_prb_constraint} \\
    & r_{k,l}^{c,t} \leq r_{max} \quad \forall c,t,k,l \in \{1,\ldots,L\}\\
    & r_{k,l}^{c,t} \leq B_C a (P_{k,l}^{c,t}E_{k,l}^{c})^b \quad \forall c,t,k,l \label{eq:rate_constraint}
\end{align}

$\mathbf{P_d}(\hat{\mathcal{U}},\mathcal{V}_d,L)$ is a convex NLP problem as long as $b \leq 1$ in \eqref{eq:rate_constraint}, which can be solved  easily offline.

For a given realization $\omega(U)$, a given $L$, and a configuration $\mathcal{V}_d$ we solve $\mathbf{P}_d(\hat{\mathcal{U}},\mathcal{V}_d,L)$ for every group $\hat{\mathcal{U}} \subseteq \{\hat{\mathcal{U}}_1,\ldots,\hat{\mathcal{U}}_{G(S_d)}\}$. This computation uses the approximated MCS function \eqref{eq:ApproxMCS}. After determining the optimal powers $\{P_{k,l}^{c,t}\}$ we compute the per-stream SINRs for all the streams in each home in the group $\hat{\mathcal{U}}$ for each PRB corresponding to subchannels in $\mathcal{C}_d(\hat{\mathcal{U}})$, i.e., $SINR_{k,l}^{c,t} = P_{k,l}^{c,t} E_{k,l}^c$ and map the result through the practical 3GPP MCS function to determine the real rate $\bar{r}_{k,l}^{c,t}$ assigned to each user stream in each PRB. Note that if $SINR_{k,l}^{c,t}<\gamma_1$, then stream $l$ of home $k$ receives a zero rate in PRB $(c,t)$, which is equivalent to not selecting stream $l$ in that PRB. The real DL rate given to each home in  $\hat{\mathcal{U}}$ in the frame is 
\begin{equation}
    \bar{\lambda}_k = \frac{1}{T} \sum_{c \in \mathcal{C}_d(\hat{\mathcal{U}})} \sum_{t \in \mathcal{T}_d}  \sum_{l=1}^{M_{\rm H}} \bar{r}_{k,l}^{c,t} \quad \forall k \in \hat{\mathcal{U}}.
\end{equation}
If $\bar{\lambda}_k \geq MBR_d$ for all $k \in \hat{\mathcal{U}}$ then $(\mathcal{V}_d,L)$ is feasible for  group $\hat{\mathcal{U}}$ of the realization $\omega(U)$ on the DL. If  $(\mathcal{V}_d,L)$ is feasible for all $\hat{\mathcal{U}} \subseteq \{\hat{\mathcal{U}}_1,\ldots,\hat{\mathcal{U}}_{G(S_d)}\}$, we say that $(\mathcal{V}_d,L)$ is feasible for realization $\omega(U)$ on the DL.

As mentioned earlier, for a given system setting $\mathcal{S}$, we consider a large set $\Omega(U)$ of realizations with $U$ homes. If there exists a pair $(\mathcal{V}_d,L)$ for which 95\% of the realizations are feasible, then we say that $(\mathcal{V}_d,L)$ is $U$-feasible on the DL. As also noted earlier, there could be many feasible such pairs, which could have an impact on the way the system is operated as discussed in Section~\ref{sect:operation}.

\subsection{Uplink Formulation}
For a given setting and radius, for a given realization $\omega(U)$,  an uplink configuration $\mathcal{V}_u=(T_u,S_u)$, the uplink user selection (i.e., user grouping) process divides the homes into static groups of size $S_u$. There is a total of $G(S_u) = \left\lceil \frac{U}{S_u} \right\rceil$ groups $\hat{\mathcal{U}}_1,\ldots,\hat{\mathcal{U}}_{G(S_u)}$, and each group $\hat{\mathcal{U}}$ is randomly assigned a set $\mathcal{C}_u(\hat{\mathcal{U}})$ of $n(S_u) = \left\lfloor \frac{C}{G(S_u)} \right\rfloor$ subchannels throughout the frame. Given $L\leq M_{\rm H}$ streams per home, the per-stream effective channels are computed for each home in the group and for each  subchannel in $\mathcal{C}_u(\hat{\mathcal{U}})$.

Given the uplink piecewise constant MCS function with a set $\mathcal{Q}$ of $Q$ levels, an UL configuration $\mathcal{V}_u$, a group $\hat{\mathcal{U}}$ and its allocated subchannels $\mathcal{C}_u(\hat{\mathcal{U}})$, knowing the effective channels $\{E_{k,l}^c\}$, $k \in \hat{\mathcal{U}}$, $l \in \{1,\ldots,L\}$, $c \in \mathcal{C}_u(\hat{\mathcal{U}})$, the objective is to verify whether there exists a distribution of power $\{P_{k,l}^{c,t}\}$ to each user stream in each PRB corresponding to channels in $\mathcal{C}_u(\hat{\mathcal{U}})$ so that we can give $MBR_u$ to all the homes in $\hat{\mathcal{U}}$. Let $\mathcal{T}_u$ be the set of UL time-slots, where $|\mathcal{T}_u| = T_u$. We directly formulate the following approximated problem  $\mathbf{P_u}(\hat{\mathcal{U}},\mathcal{V}_u,L)$ on the UL (similar to $\mathbf{P_d}(\hat{\mathcal{U}},\mathcal{V}_d,L)$ on the DL) as:
\begin{align}
  & \max_{\lambda'(\hat{\mathcal{U}},\mathcal{V}_u,L),(P_{k,l}^{c,t}), (r_{k,l}^{c,t}), (\lambda_k')} \lambda'(\hat{\mathcal{U}},\mathcal{V}_u,L) \\
    & \lambda_k' \geq \lambda'(\hat{\mathcal{U}},\mathcal{V}_u,L) \quad \forall k \in \hat{\mathcal{U}} \\
    & \lambda_k' = \frac{1}{T} \sum_{c \in \mathcal{C}_u(\hat{\mathcal{U}})} \sum_{t \in \mathcal{T}_u}  \sum_{l=1}^{L} r_{k,l}^{c,t} \quad \forall k \in \hat{\mathcal{U}}\\
    & \sum_{l=1}^{L} P_{k,l}^{c,t} \leq P_{PRB}^k = \frac{P_{\rm H}}{|\mathcal{C}_u(\hat{\mathcal{U}})|} \quad \forall c \in \mathcal{C}_u(\hat{\mathcal{U}}), t \in \mathcal{T}_u,k
    \label{eq:ul_prb_constraint}\\
    & r_{k,l}^{c,t} \leq r_{max} \quad \forall c,t, k, l \\
    & r_{k,l}^{c,t} \leq B_C a (P_{k,l}^{c,t} E_{k,l}^{c})^b  \quad \forall c,t, k, l
\end{align}

The key difference in the UL problem is seen in \eqref{eq:ul_prb_constraint} where each home has its own power budget to distribute over its allocated channels, instead of the  power budget of the BS in the DL that has to be across all channels in~\eqref{eq:dl_prb_constraint}.

For the set of realizations $\Omega(U)$ we similarly define the notion of $U$-feasibility on the uplink for a pair $(\mathcal{V}_u,L)$. Finally, we say that a pair $(\mathcal{V},L)$ is $U$-feasible for the system if $(\mathcal{V}_d,L)$ is $U$-feasible on the DL \emph{and} $(\mathcal{V}_u,L)$ is $U$-feasible on the UL.

\subsection{Computing User Limits}\label{sect:UserLimit}
We aim to compute the user limit $U^*$ for a given setting $\mathcal{S}$ and a given radius $\mathcal{R}$. 
Let $\mathcal{F}(\mathcal{V},L)$ be the set of $U$ for which $(\mathcal{V},L)$ is $U$-feasible, for the system; then
\begin{equation}
  U^*=\max_{T_u,S_d,S_u,L} \{U \in \mathcal{F}(\mathcal{V},L)\}.  
\end{equation}

Computing $U^*$ using brute force is very cumbersome since  we need to solve, for increasing values of $U$,  and many realizations, problems $\mathbf{P_d}(\hat{\mathcal{U}},\mathcal{V}_d,L)$ and  $\mathbf{P_u}(\hat{\mathcal{U}},\mathcal{V}_u,L)$ for each quadruple $(T_u, S_d, S_u,L)$ ($1\leq T_u \leq T-1$, $1\leq S_d,S_u\leq N$, $1\leq L\leq M_{\rm H}$) for all groups created on the DL and the UL by the DL and UL user selection.  We can stop when for a given value of $U$, no quadruple is $U$-feasible. However, the value of $U^*$ might be large if the number $C$ of subchannels in the system is large and the radius is not too large making this computation time-consuming.

Instead, we propose a much simpler way to compute $U^*$. For a given system setting $\mathcal{S}$, a configuration $\mathcal{V} = (T_u,S_d,S_u)$, $L$ streams,  we compute for the DL (resp. the UL), how many subchannels $n(\mathcal{V}_d,L)$ (resp. $n(\mathcal{V}_u,L)$) are needed to offer $MBR_d$ on the DL (resp. $MBR_u$ on the UL) to 95\% of the realizations, made of  $S_d$ (resp. $S_u$) homes, in a set  $\Omega_d$ (resp. $\Omega_u$). Since $S_d \leq N$ and $S_u \leq N$, the amount of computations is much lower than with the brute force approach.

Specifically, given a setting $\mathcal{S}$ characterized in particular by the total number of subchannels in the system $C$, a radius $\mathcal{R}$,  to compute the number of subchannels  $n(\mathcal{V}_d,L)$ (resp. $n(\mathcal{V}_u,L)$) for a given configuration $\mathcal{V}_d$ (resp. $\mathcal{V}_u$) and $L$ streams, we generate a set $\Omega_d$ of realizations (aka groups) of size $S_d$ for a given DL configuration $(T_u,S_d)$ and a set $\Omega_u$ of  realizations (aka groups) of size $S_u$ for a given UL configuration $(T_u,S_u)$ (they are not the same realizations because, in general, $S_d\neq S_u$, although $|\Omega_d|=|\Omega_u|=\Omega$).   

Focusing on the DL, given a realization $\omega_d$  with $S_d$ homes, we use Algorithm~\ref{alg:ComputeLimit} to compute $n(\mathcal{V}_d,L,\omega_d)$ the minimum number of channels necessary to offer $MBR_d$  to all $S_d$  homes in $\omega_d$  given $T_u$ time-slots are allocated to the UL. Note that for a given MBR, it is impossible to give a home the MBR if the number of subchannels allocated to a home is less than $C_{\rm min}(MBR) = \left\lceil \frac{MBR}{B_C L SE_{\rm max}} \right\rceil$, where $SE_{max}$ is the highest SE of the practical MCS function. In other words, Line \ref{line:checkfeasible} will always be false when $n < C_{\rm min}(MBR)$, hence we initialize it with $n = C_{\rm min}(MBR)$ in Line~\ref{line:MinSubchannels}. There is a subtlety in Line~\ref{line:SolveOpt} for the DL. Indeed, on the DL, we solve the PD problem with $n$ subchannels allocated to the group but with $P_{PRB}^d= \frac{P_{\rm max}}{C}$ where $C$ is the number of subchannels allocated to the original system; hence $n(\mathcal{V}_{d},L)$ is a function of $C$. 

\begin{algorithm}[t]
    \caption{Compute $n(\mathcal{V}_{d/u},L,\omega_{d/u})$}
    \label{alg:ComputeLimit}
    \begin{algorithmic}[1]
        \State $n \leftarrow C_{\rm min}(MBR_{d/u}) = \left\lceil \frac{MBR_{d/u}}{B_C L SE_{\rm max}}  \right\rceil$ \label{line:MinSubchannels}
        \State Solve $\mathbf{P_{d/u}}(\omega_{d/u},\mathcal{V}_{d/u},L)$ with $n$ subchannels \label{line:SolveOpt}
        \If{all $S_{d/u}$ homes receive $MBR_{d/u}$} \label{line:checkfeasible}
            \State $n(\mathcal{V}_{d/u},L,\omega_{d/u}) \leftarrow n$
        \Else
            \State $n \leftarrow n + 1$
            \If{$n \leq C$}
                \State Go to \ref{line:SolveOpt}
            \Else                
                \State $\omega_{d/u}$ infeasible
            \EndIf
        \EndIf
        \end{algorithmic}
\end{algorithm}

For the UL, we also use Algorithm~\ref{alg:ComputeLimit} but when we solve the problem with $n$ subchannels, $P_{PRB}^k = \frac{P_{\rm H}}{n}$.

Having computed $n(\mathcal{V}_d,L)$ (resp. $n(\mathcal{V}_u,L)$), we can compute the number of homes that can receive $MBR_d$ on the DL (resp. $MBR_u$ on the UL) if the system has $C$ subchannels, i.e., 
\begin{align}
    U_d(\mathcal{V}_d,L,C) &= \left\lfloor\frac{C}{n(\mathcal{V}_d,L)}\right\rfloor S_d \label{eq:dlLimit} \\
    U_u(\mathcal{V}_u,L,C) &= \left\lfloor\frac{C}{n(\mathcal{V}_u,L)}\right\rfloor S_u \label{eq:ulLimit}
\end{align}
This is because we can create independent groups of size $S_d$ and $S_u$ respectively and allocate to each group the number of subchannels needed.
Note that the above equations are true because when we allocate subchannels to groups the power per PRB neither changes on the DL (we keep $P_{PRB}^d=\frac{P_{\rm max}}{C}$ irrespective of the value of $S_d$) nor does it change on the UL (by construction). \eqref{eq:dlLimit} and \eqref{eq:ulLimit} could be computed exhaustively for every quadruple $(T_u,S_d,S_u,L)$, but more efficient techniques can be applied to reduce computations (e.g., if $(T_u,S_d,L)$ is determined to be infeasible in the DL, then we know that $(\widetilde{T}_u,S_d,L)$ is infeasible in the DL for all $\widetilde{T}_u > T_u$).

A search on $T_u$ for different $S_d$ and $S_u$ then determines the correct configuration of $T_u$, $S_d$, $S_u$ required to maximize the number of homes in the system. 
Let $U^*(L,C)$ be the user limit for the system setting $\mathcal{S}$ with $C$ subchannels when there are $L$ streams per home, where
\begin{equation}\label{eq:AbsLimit}
    U^*(L,C) = \max_{T_u,S_d,S_u} \left\{ \min (U_d(\mathcal{V}_d,L,C), U_u(\mathcal{V}_u,L,C)) \right\}.
\end{equation}
To keep the notation light, we omit $(L,C)$ in further references to the user limit. $U^*$ is obtained for some $T_u = T_u^*$, $S_d= S_d^*$, and $S_u= S_u^*$. We refer to the triple $\mathcal{V}^* = (T_u^*,S_d^*,S_u^*)$ yielding $U^*$ as the \emph{optimal configuration}.

\begin{table*}[t!]
\centering
\caption{System Parameters for Numerical Results}
\begin{tabular}{l|l}
\hline
\textbf{Parameters}  & \textbf{Values} \\
\hline
Carrier frequency ($f_c$) & 3.5\,$\rm GHz$ \\
Cell radius ($\mathcal{R}$)  & \{1500,2000,3000,4000,5000,10000\}\,$\rm m$ \\
System bandwidth ($B$) & \{25,50\}\,MHz\\
Number of sub-channels ($C$) & \{65,133\}\\
Subchannel bandwidth ($B_C$) & 360\,$\rm kHz$\\
Number of time-slots per frame ($T$) & 20 \\
Time-slot duration ($T_S$) & 0.5\,$\rm ms$\\
Maximum power of BS ($P_{\rm max}$) & \{40,80\}\,W  \\
Maximum power of home ($P_{\rm H}$) & 400\,mW  \\
Total antennas at BS ($M_{\rm BS}$) & \{64,128\} \\
Total antennas at home ($M_{\rm H}$) & \{4,8\} \\ 
Minimum bit-rate in DL ($MBR_d$) & \{15,30\}\,Mbps \\
Minimum bit-rate in UL ($MBR_u$) & \{2.5,5\}\,Mbps \\
Minimum distance between BS and home ($d_{2D}^{\min}$)  & 35\,$\rm m$ \\
Effective antenna height at BS ($h_{\rm BS}$) & 100\,m \\
Effective antenna height at home ($h_{\rm H}$) & 5\,m \\
Adjacent antenna distance at BS ($d_a^{\rm BS}$) & 0.5\,$\rm \lambda$ \\
Adjacent antenna distance at home ($d_a^{\rm H}$) & 0.2\,m \\
BS channel correlation coefficient ($\varphi_{\rm BS}$) & $0.4$ \\
Home channel correlation coefficient ($\varphi_{\rm H}$) & $\exp\{-d_a^{\rm H}/\lambda\}=0.01$  \\
Noise power density ($N_0$) & --174\,$\rm dBm/Hz$ \\
Noise figure ($\zeta$) & 9\,dB \\
Rician factor ($\kappa$) & $\mathcal{N}(7,4^2)$\,dB \\
\hline
\end{tabular}
\label{Tab:SimulationParameters}
\end{table*}

Clearly this simplified process gives us an estimate of what we could find through brute force (we will validate it in the numerical section). If necessary, we could use this estimate as a starting point to initialize a search for the real limit. The fact that the parameters in the configuration $(T_u,S_d,S_u)$ are all integer values makes our search space relatively small; however, this also introduces the possibility that rounding errors could occur when using our approach. That being said, in Section~\ref{sect:validation} we will validate that the granularity provided by the simplified process we developed is acceptable for the purpose of planning a network. The user limit that we obtain also depends on our grouping strategy. In practice, for a given situation, i.e., a set of user locations and their channels, an operator might benefit from grouping users more smartly; however, because we are dealing with many realizations of users we resort to arbitrary grouping, leaving smarter grouping for further study in an operational context as opposed to a planning one. It is also possible that a more sophisticated user selection could offer the MBRs to more homes if we were to reformulate the problem as a \emph{joint} user selection and power distribution problem that was solved per-realization. However, this would make the problem more difficult to compute and it could only be solved through brute force. The strength of our approach is that it is simple and robust and provides very useful insights to MNOs during the planning phase.

\section{Numerical Results}\label{sect:results}

In this section we present the numerical results of our study. We begin by describing the channel settings. Next we conduct an initial study of the case when $U \leq N$ and all $U$ homes are selected in each PRB, which informs us about the impact of multiple streams on performance as well as how we can implement user grouping. Then we study the network planning problem when $U > N$, determining the user limit and the optimal configuration of $T_u$, $S_d$, and $S_u$ for a given system setting and radius. We end the section by providing some useful insights into how MNOs could operate FWA networks.

\subsection{System and Channel Settings}\label{sect:channel}

\begin{table*}[t!]
\centering
\caption{Tapped-Delay Line and Power Delay Profile~\cite{3gppTR38901}}\label{Table:TapsDelay}
\setlength{\tabcolsep}{1mm}{
\begin{tabular}{|c|c|c|c|c|c|c|c|c|}
\hline
Tap number & 1 & 2 & 3 & 4 & 5 & 6 & 7 & 8 \\
\hline
Delay $t_d$ (ns) & 0 & 51.33 & 54.40 & 56.30 & 54.40 & 71.12 & 190.92 & 192.93 \\
\hline
$\varrho_d$ & 0.9209 & 0.0244 & 0.0144 & 0.0097 & 0.0048 & 0.0053 & 0.0128 & 0.0077 \\
\hline
\end{tabular}}
\end{table*}

\begin{table*}[t]
    \centering
    \caption{SINR-SE Mapping Used for Practical MCS~\cite{zhang2021joint,lopez2011optimization}}
    \begin{tabular}{|c|c|c|c|c|c|c|c|c|c|c|c|c|c|c|c|}
        \hline
        SINR (dB) & -6.5 & -4 & -2.6 & -1 & 1 & 3 & 6.6 & 10 & 11.4 & 11.8 & 13 & 13.8 & 15.6 & 16.8 & 17.6 \\
        \hline
        Spectral efficiency (bps/Hz) & 0.14 & 0.22 & 0.36 & 0.56 & 0.82 & 1.10 & 1.38 & 1.78 & 2.25 & 2.55 & 3.10 & 3.64 & 4.22 & 4.78 & 5.18 \\
        \hline
    \end{tabular}
    \label{tab:mcs}
\end{table*}

Before providing the numerical results, we describe the channel settings and then the MCS function. We adopt the 3GPP rural macro (RMa) path loss model \cite{3gppTR38901}, assume that the MNO installs a roof-mounted array of antennas at each home and points it directly at the best BS, guaranteeing a line-of-sight (LoS) view between every home and the BS. The BS transmits using a mid-band carrier frequency $f_c = 3.5$\,GHz. We consider NR numerology 1 \cite{3gppTS38211}, which utilizes subchannels of bandwidth $B_C = 360$\,kHz and time-slots of length $T_S = 0.5$\,ms. The frame has a total of $T = 20$ time-slots. We will vary the number of slots allocated to the uplink $T_u$ in order to determine the user limit for a given system setting $\mathcal{S}$. We consider system bandwidths of $B = 25$\,MHz and $B = 50$\,MHz, which correspond to $C = 65$ and $C = 133$ subchannels, respectively\footnote{Due to the use of guard bands, the total number of subchannels does not scale linearly with the bandwidth.}~\cite{3gppTS381011}.

The BS is located at the center of the cell and the homes are uniformly distributed within a circle of radius $\mathcal{R}$ and a minimum distance of $d_{2D}^{\min} = 35$\,m from the BS. The BS antennas are at height $h_{\rm BS} = 100$\,m and the home antennas are at height $h_{\rm H} = 5$\,m, which is considered a typical building height under the RMa model.

We will consider several settings $\mathcal{S}$, by varying the number of antennas at the BS $M_{\rm BS}$, the number of antennas at each home $M_{\rm H}$, the BS transmit power $P_{\rm max}$, the system bandwidth $B$ (and hence, the total subchannels $C$), and the MBR in the DL $MBR_d$ and UL $MBR_u$ for different cell radii $\mathcal{R}$. We summarize the parameters used in the numerical computations in Table~\ref{Tab:SimulationParameters}. The system settings that we vary are shown in braces \{\}. 

Since each home is fixed, the channel coherence time is large; hence, we assume that the channel coefficients are constant throughout the frame. A tapped-delay-line channel model with $D$ taps is adopted, where the channel matrix $\mathbf{G}_k^{c,t}$ from the BS to the \mbox{$k$-th} home at PRB~$(c,t)$ can be written with respect to the $D$ channel taps as~\cite{liu2008cautious}:
\begin{equation}\label{Eq:FrequencyChannelMatrix}
    \mathbf{G}_k^{c,t} = \mathbf{G}_k^c = \sum_{d = 1}^{D} \overline{\mathbf{G}}_{k}[d] \exp{\left(-j 2 \pi \tau_d c / C \right)},
\end{equation}
where $\overline{\mathbf{G}}_{k}[d]$ is the channel matrix of the \mbox{$d$-th} tap, which is constant throughout the frame and has a normalized delay of $\tau_d$. Note that the delay of the \mbox{$d$-th} tap $t_d = \tau_d / B$, where $B$ is the system bandwidth. In our system, we utilize an $8$-tap channel model, where the value of $t_d$ is given in Table~\ref{Table:TapsDelay}~\cite{3gppTR38901}.

With a LoS component from the BS, the first channel tap of each home follows a Rician distribution, i.e.,
\begin{equation}
    \overline{\mathbf{G}}_{k}[1] = \sqrt{ \varrho_{1} \beta_{k}} \left( \sqrt{\frac{1}{1+\kappa_k}} \mathbf{R}_{\rm H}^{\frac{1}{2}} \mathbf{H}_G[1] \mathbf{R}_{\rm BS}^{\frac{1}{2}}+  \sqrt{\frac{\kappa_{k}}{1+\kappa_{k}}} \mathbf{H}_{k} \right),
\end{equation}
where $\kappa_{k}$ is the Rician factor and $\mathbf{H}_{k} \in \mathbb{C}^{M_{\rm H} \times M_{\rm BS}}$ represents the LoS component from the BS to the \mbox{$k$-th} home. $\beta_{k}$ denotes the \mbox{large-scale} fading channel coefficient, involving the power attenuation and shadowing between the BS and the \mbox{$k$-th} home, which is determined from the 3GPP RMa 3D path loss model shown in Table 7.4.1-1 of~\cite{3gppTR38901}. $\mathbf{H}_G[d] \in \mathbb{C}^{M_{\rm H} \times M_{\rm BS}}$, $d = 1,2,\ldots,D$, is a random matrix whose entries are \mbox{zero-mean} i.i.d complex Gaussian random variables with unit variance; and $\mathbf{R}_{\rm H}$ (resp. $\mathbf{R}_{\rm BS}$) is the correlation matrix at the home (resp. the BS) assumed to be the same for every home and to not change with time. $\varrho_{d}$ represents the normalized power of the \mbox{$d$-th} tap, and its value is shown in Table~\ref{Table:TapsDelay}. In terms of $\mathbf{H}_k$, the LoS component from the \mbox{$m$-th} antenna of the BS, located at $\mathbf{a}_m \in \mathbb{R}^3$, to the \mbox{$n$-th} antenna of the \mbox{$k$-th} home, located at $\mathbf{u}_k^n \in \mathbb{R}^3$, is given by
\begin{equation}
    h_k^{m,n} = e^{ 2 \pi j \parallel \mathbf{u}_k^n - \mathbf{a}_m\parallel/\lambda },
\end{equation}
where $\lambda = 3 \times 10^8/f_c$ is the wavelength, and $f_c$ is the carrier frequency. All other channel taps follow a Rayleigh distribution~\cite{3gppTR38901}, i.e.,
\begin{equation}\label{Eq:RayleighChannel}
    \overline{\mathbf{G}}_{k}[d] = \sqrt{\varrho_{d} \beta_{k}}  \mathbf{R}_{\rm H}^{\frac{1}{2}} \mathbf{H}_G[d] \mathbf{R}_{\rm BS}^{\frac{1}{2}}, \quad d = 2,3,\ldots, D.
\end{equation}

The BS and all homes are assumed to be equipped with uniform planar arrays (UPA) with omnidirectional antennas; thus at the BS (resp. homes) there are $\sqrt{M_{\rm BS}}$ (resp. $\sqrt{M_{\rm H}}$) antennas in both vertical and horizontal directions with identical spacing $d_a^{\rm BS}$ (resp. $d_a^{\rm H}$). Following~\cite{zhang2021joint,ying2014kronecker,lim2017bounds}, we adopt the exponential correlation model, where the \mbox{$(i,j)$-th} entry $[\mathbf{R}]_{i,j}$ of the correlation matrix $\mathbf{R}$ is given by
\begin{equation}
    [\mathbf{R}_{\rm BS}]_{i,j} = \varphi_{\rm BS}^{|v(i) - v(j)| + |h(i) - h(j)|}, \,\, i,j = 1,2,\ldots,M_{\rm BS},
\end{equation}
\begin{equation}
    [\mathbf{R}_{\rm H}]_{i,j} = \varphi_{\rm H}^{|v(i) - v(j)| + |h(i) - h(j)|}, \quad i,j = 1,2,\ldots,M_{\rm H},
\end{equation}
where $0 \leq \varphi_{\rm BS}, \varphi_{\rm H} \leq 1$ denote the correlation coefficients between adjacent antennas at the BS and between adjacent antennas at homes, respectively; $v(m)$ and $h(m)$ are the vertical and horizontal indexes of the \mbox{$m$-th} antenna of the UPA. For the BS we set $\varphi_{\rm BS} = 0.4$~\cite{zhang2021joint}. For the homes, since the number of FWA antennas is assumed to be small, the antenna separation $d_a^{\rm H}$ is large relative to $\lambda$ and we set $\varphi_{\rm H} = e^{-d_a^{\rm H}/\lambda}=0.01$~\cite{fan2013dual}.

NR defines several MCS tables for the DL and UL~\cite{3gppTS38214}, and the network and user equipment can switch from one table to another depending on the radio conditions. In general, the relationship between the radio channel and the selected MCS is not standardized and is typically derived through system and/or link-level simulations \cite{lagen2020new,ikuno2010system}. For simplicity we use only one of the NR MCS tables in our system. The table we use supports up-to 64-QAM modulation and is the table that would typically be selected for average-to-poor channel conditions (i.e., for cell edge users in a rural environment). In NR this table can be used for the DL and the UL (in LTE it was only used for DL) so we use it for both. The corresponding SINR-SE mapping is shown in Table~\ref{tab:mcs}~\cite{zhang2021joint,lopez2011optimization}.

For the MCS approximation in \eqref{eq:ApproxMCS} we follow the results derived in~\cite{zhang2021joint} to approximate Table~\ref{tab:mcs} and set $a = 0.648$ and $b = 0.5$. Thus, the rate of the \mbox{$l$-th} stream of home $k$ in PRB~$(c,t)$ is approximated by 
\begin{equation}
    r_{k,l}^{c,t} = B_C \min (SE_{\rm max},0.648(P_{k,l}^{c,t} E_{k,l}^c)^{0.5})
\end{equation}

\subsection{Understanding the Impact of Streams and Group Sizes}
\label{toto}

\begin{figure*}[t]
    \centering
        \begin{minipage}{0.32\linewidth}
        \centering
        \includegraphics[width=0.96\textwidth]{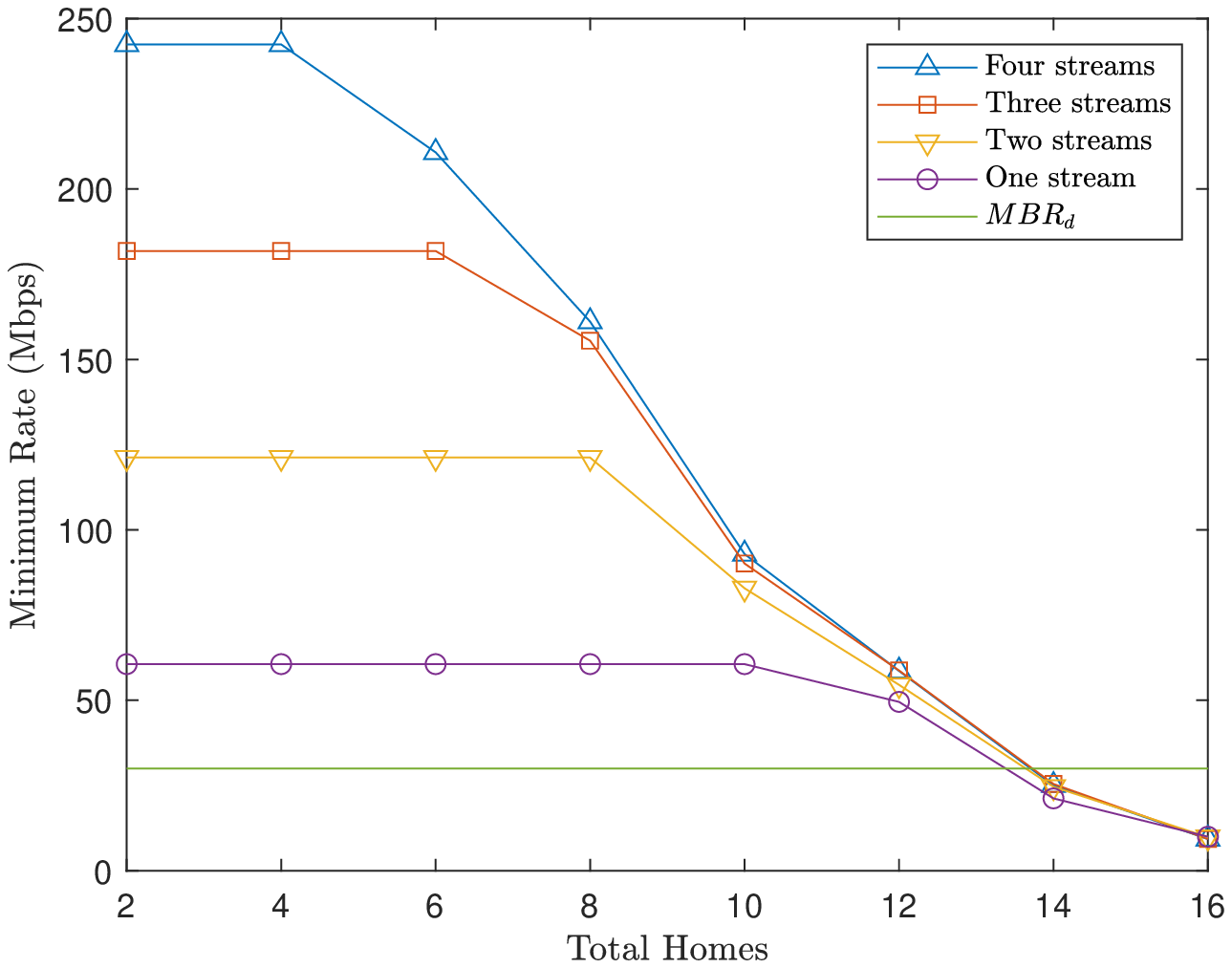}
        \subcaption{$C = 65$ total subchannels in system.}
        \label{fig:min_rate_streams_DL_65ch}
    \end{minipage}
    \begin{minipage}{0.32\linewidth}
        \centering
        \includegraphics[width=0.96\textwidth]{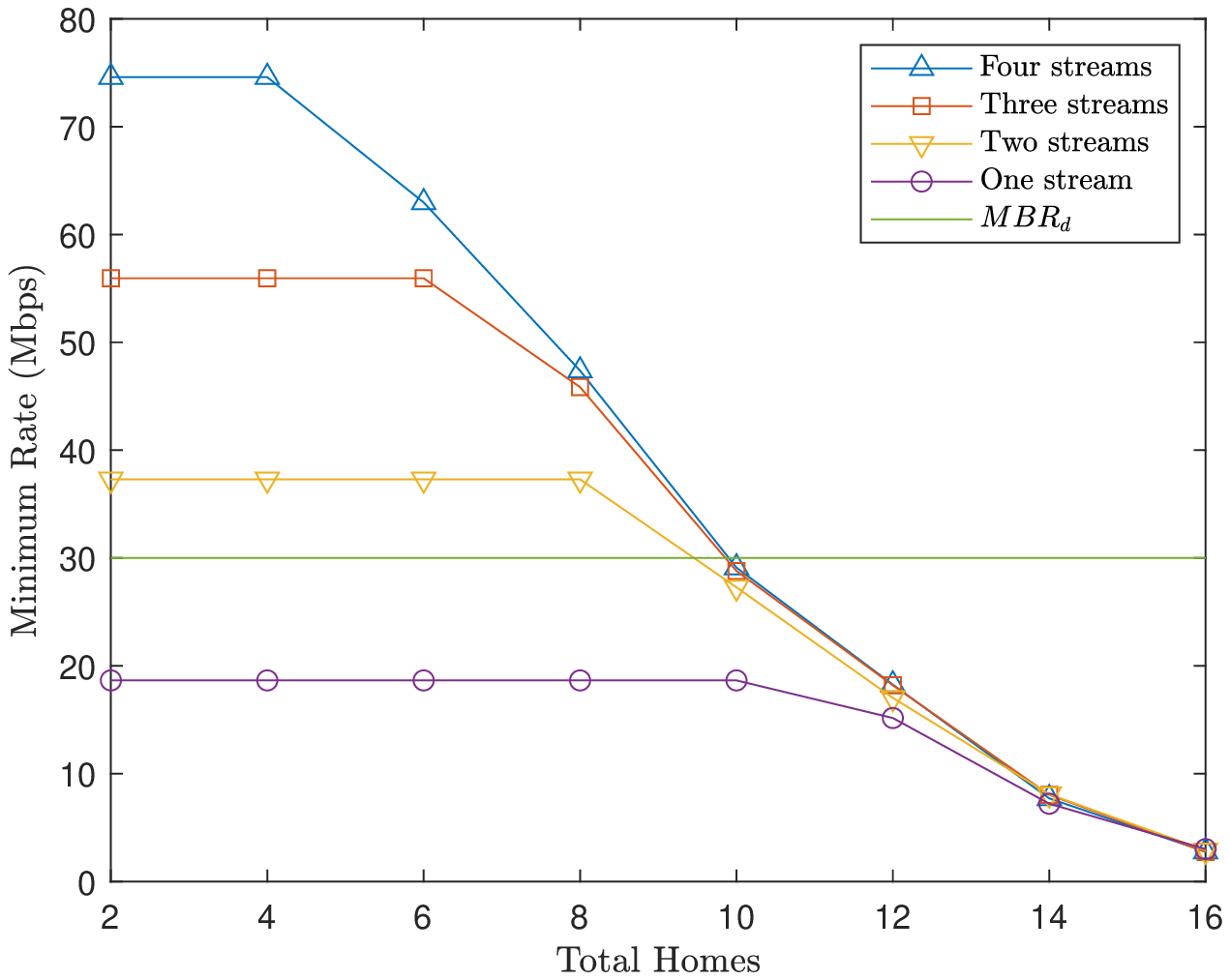}
        \subcaption{$C = 20$ total subchannels in system.}
        \label{fig:min_rate_streams_DL_20ch}
    \end{minipage}
    \begin{minipage}{0.32\linewidth}
        \centering
        \includegraphics[width=0.96\textwidth]{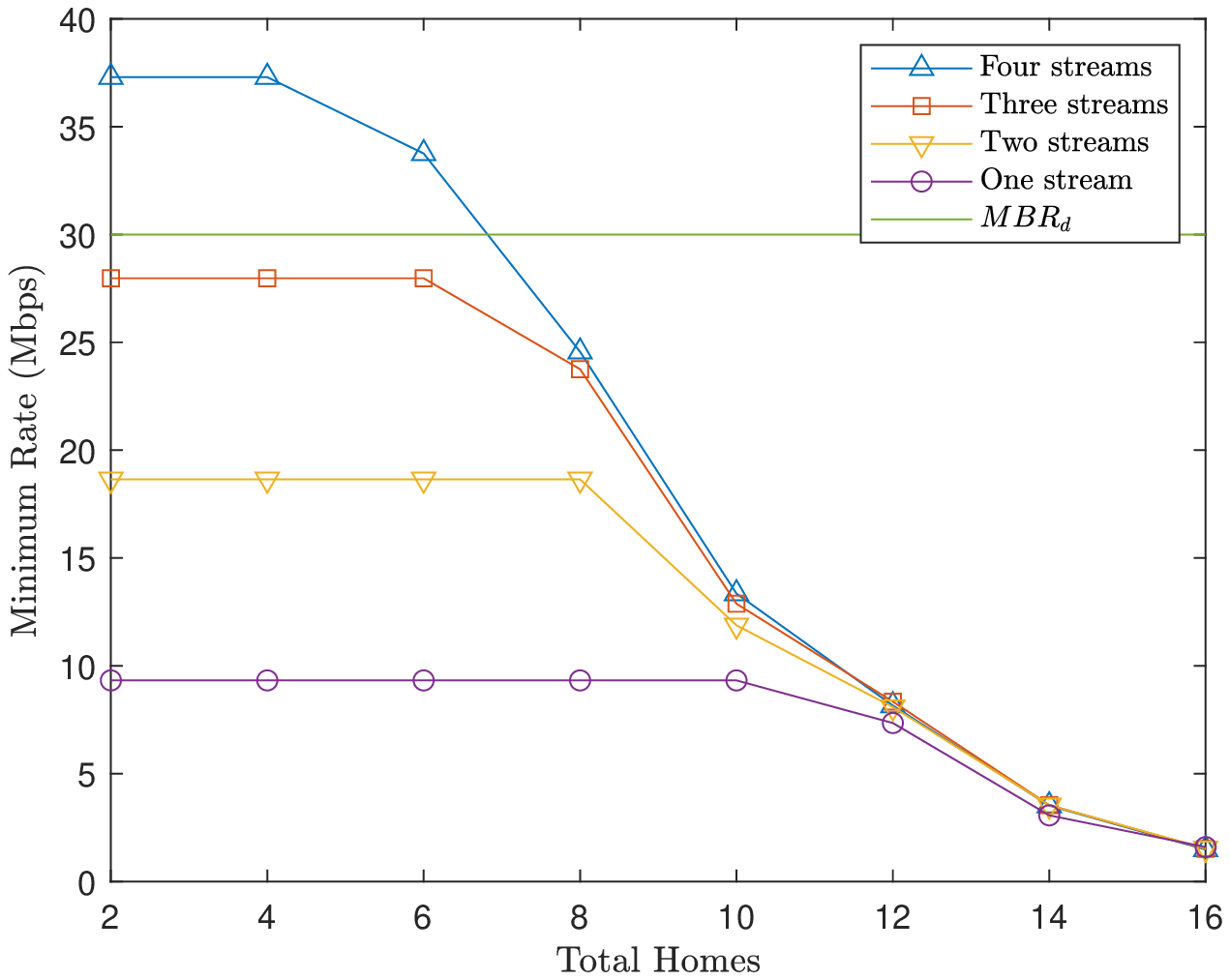}
        \subcaption{$C = 10$ total subchannels in system.}
        \label{fig:min_rate_streams_DL_10ch}
    \end{minipage}
    \caption{$95^{th}$ percentile of minimum DL user rates for $U \leq N$: \mbox{$\Omega = 100$}, \mbox{$\mathcal{R} = 1500$\,m}, \mbox{$B = 25$}\,MHz, \mbox{$M_{\rm BS} = 64$}, \mbox{$M_{\rm H} = 4$}, \mbox{$P_{\rm max} = 40$\,W}, \mbox{$MBR_d = 30$\,Mbps, $T_u = 10$}. The power per PRB is $\frac{P_{\rm max}}{65}$ for all cases.}
    \label{fig:min_rate_streams_DL}
\end{figure*}
\begin{figure*}[t]
    \centering
        \begin{minipage}{0.32\linewidth}
        \centering
        \includegraphics[width=0.96\textwidth]{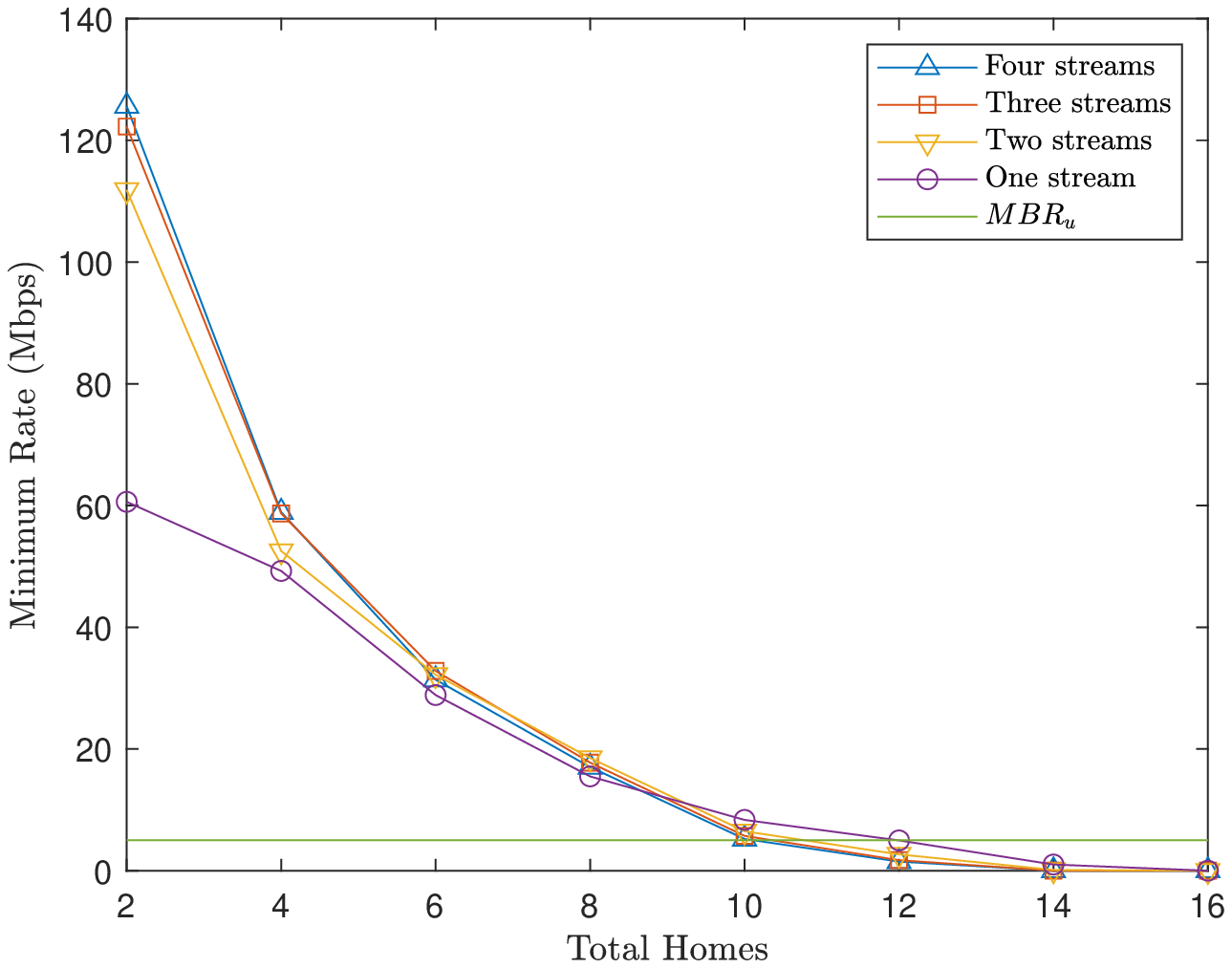}
        \subcaption{$C = 65$ total subchannels in system.}
        \label{fig:min_rate_streams_UL_65ch}
    \end{minipage}
    \begin{minipage}{0.32\linewidth}
        \centering
        \includegraphics[width=0.96\textwidth]{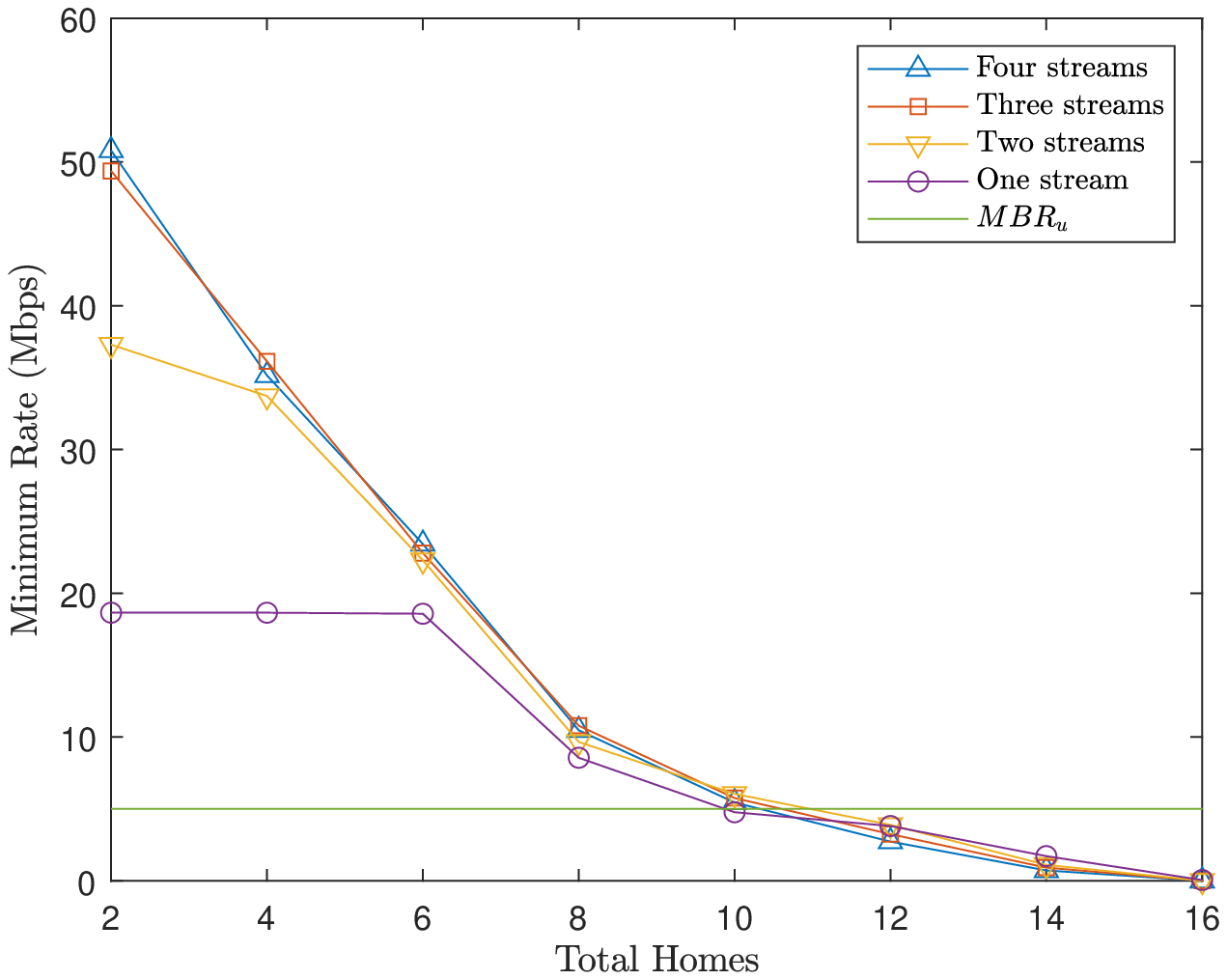}
        \subcaption{$C = 20$ total subchannels in system.}
        \label{fig:min_rate_streams_UL_20ch}
    \end{minipage}
    \begin{minipage}{0.32\linewidth}
        \centering
        \includegraphics[width=0.96\textwidth]{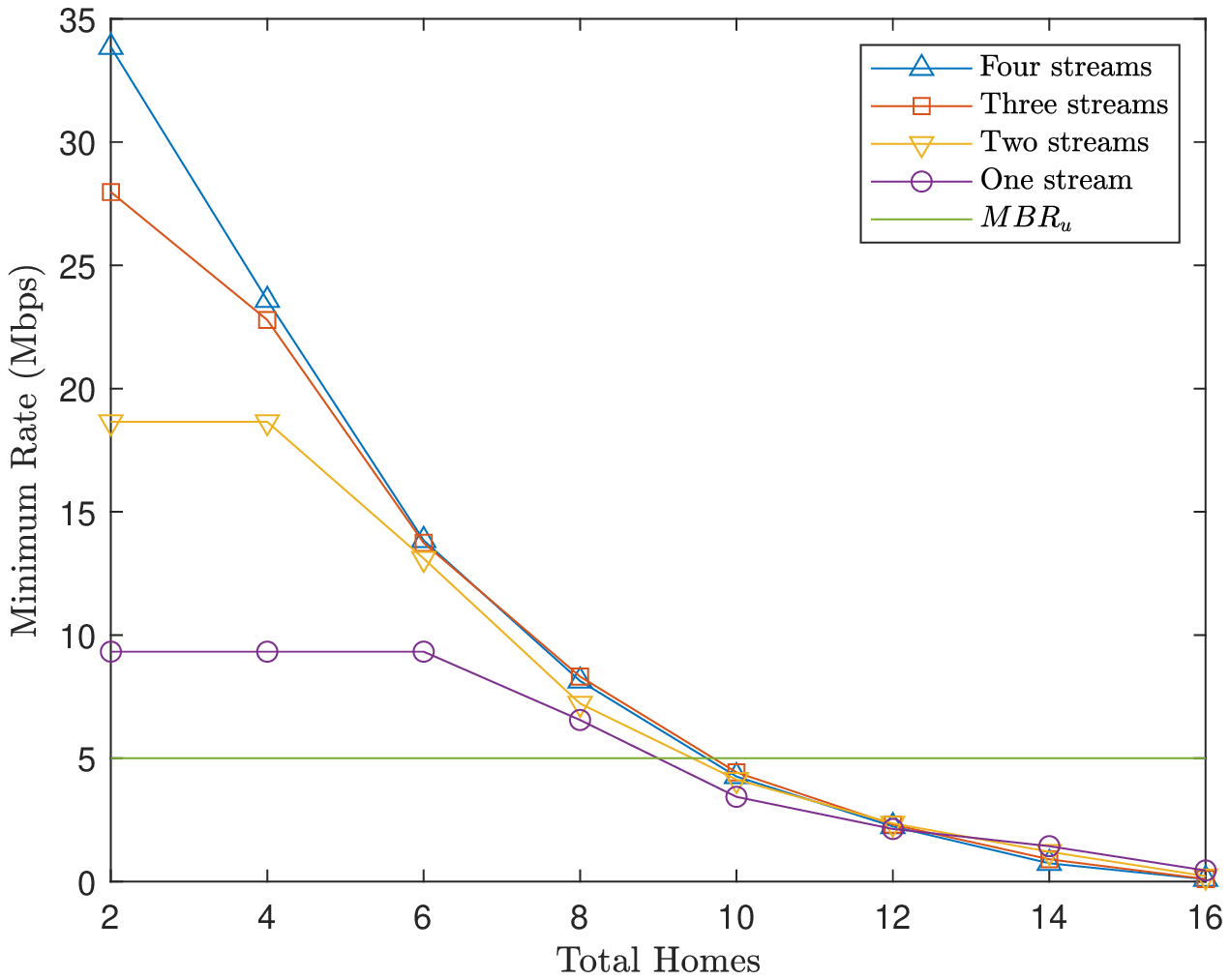}
        \subcaption{$C = 10$ total subchannels in system.}
        \label{fig:min_rate_streams_UL_10ch}
    \end{minipage}
    \caption{$95^{th}$ percentile of minimum UL user rates for $U \leq N$: \mbox{$\Omega = 100$}, \mbox{$\mathcal{R} = 1500$\,m}, \mbox{$B = 25$}\,MHz, \mbox{$M_{\rm BS} = 64$}, \mbox{$M_{\rm H} = 4$}, \mbox{$P_{\rm H} = 400$\,mW}, \mbox{$MBR_u = 5$\,Mbps, $T_u = 10$}.}
    \label{fig:min_rate_streams_UL}
\end{figure*}

As a first step, we study the case where $U \leq N$ for $L \in \{1,\ldots,M_{\rm H}\}$ and all $U$ homes are selected in each PRB, i.e., there is a single group of size $U$. This will provide insight into how user grouping should be performed when $U > N$ (basically, what we will show is that using $S_d = N = S_u$ is \emph{not} the right thing to do when $U > N$). We will also observe the impact  on the performance of limiting the number of  streams to less than $M_{\rm H}$.

In this first numerical section devoted to $U\leq N$, we set \mbox{$B = 25$\,MHz}, \mbox{$M_{\rm BS} = 64$  antennas}, \mbox{$M_{\rm H} = 4$ antennas}, \mbox{$P_{\rm max} = 40$\, W}, \mbox{$MBR_d = 30$\, Mbps} and \mbox{$MBR_u = 5$\, Mbps},  which we refer to as the \emph{baseline} setting in the remainder of the paper. We will study the impact of all the above parameters later. With $M_{\rm BS} = 64$ and $M_{\rm H} = 4$ we have $N = 16$; hence we examine the case where $U \leq 16$. Also, the maximum number of streams per home is $M_{\rm H} = 4$.  We set $T_u = \frac{T}{2} = 10$ slots. Clearly, the results are very dependent on the value of $T_u$ as will be discussed at the end of this section.

Note that  our baseline setting presented throughout the rest of the paper will comprise \mbox{$C = 65$ subchannels}; however, in this subsection we change the total number of subchannels in the system $C$, while keeping the DL power per PRB fixed at $P_{\rm max} / 65$. Here, changing $C$ will reveal how the number of subchannels allocated to a group impacts the minimum rates given to homes in that group, and therefore motivates the computation of the minimum subchannels $n$ needed to provide the MBR in Algorithm~\ref{alg:ComputeLimit}.

For every $U\leq N$, we generate $\Omega=100$ realizations. Each realization $\omega(U)$ corresponds to a set $\mathcal{U}$ of $U$ homes  along with their channel matrices. For each $\omega(U)$ and each $L \leq M_{\rm H}$ we compute the effective channels and solve $\mathbf{P_d}(\mathcal{U},\mathcal{V}_d,L)$ to obtain the power distribution that maximizes the minimum DL rate of the homes, and verify using the real MCS function whether every home receives a rate of at least $MBR_d$. Likewise, we solve $\mathbf{P_u}(\mathcal{U},\mathcal{V}_u,L)$ in the UL to confirm whether each home receives a rate of at least $MBR_u$.

In Figure~\ref{fig:min_rate_streams_DL}, we plot the $95^{th}$ percentile of the minimum DL user rate (among the $U$ homes) for different numbers of streams under the baseline system settings when the radius is $\mathcal{R} = 1500$\,m  and $T_u = 10$ as a function of $U$. The results in Figures~\ref{fig:min_rate_streams_DL_65ch}, \ref{fig:min_rate_streams_DL_20ch} and \ref{fig:min_rate_streams_DL_10ch}   correspond to the cases where  the total subchannels $C$ in the system are 65, 20, or 10, respectively. For all cases the DL PRB power is fixed, i.e.,  $P_{PRB}^d = \frac{P_{\rm max}}{65}$. We observe that limiting the number of streams to less than $M_{\rm H}$ is impacting the performance a lot. Indeed, multiple streams can greatly improve the performance when $U$, the number of homes, is small relative to $N$. However, as $U$ increases, we begin to see the curves for different numbers of streams overlapping, which indicates that PD is not selecting all streams, e.g., when $U = 8$, the fourth stream provides almost no benefit.  On the other hand, the results also make it clear that for certain numbers of homes, using too few streams could greatly degrade performance, e.g., for $U < 10$, limiting the number of streams to one per home could degrade performance by a factor of 3 to 4. Therefore, it is clear that the appropriate number of streams is dependent on the selected number of homes. By using PD to perform stream selection, we can automatically determine that number.

Figure~\ref{fig:min_rate_streams_DL_65ch} also shows that with \mbox{$MBR_d = 30$ Mbps} and $T_u = 10$,  we cannot guarantee $MBR_d$ to every home for $U = 14$ or $U = 16$ even if they are allocated all the 65 subchannels. On the other hand, Figure~\ref{fig:min_rate_streams_DL_20ch} shows that with 20 subchannels we can give 8 homes $MBR_d$; hence, to give 16 homes $MBR_d$ we could simply divide the homes into two groups and give each group 20 subchannels. Thus, we learn that for $U$ close to $N$, it might be better to divide homes into groups and allocate a fraction of the subchannels to each group than to select all homes at once and give them all of the subchannels.

These results also illustrate how grouping can be used to serve $U > N$ homes. For example, continuing from above, after allocating 40 subchannels across the first two groups of 8 homes, there would be 25 unallocated subchannels remaining. Hence an additional 8 homes could easily be given $MBR_d$ for a total of 24 homes served. However, we are not limited to 24 homes: Figure~\ref{fig:min_rate_streams_DL_10ch} shows that with 10 subchannels, 6 homes can be given $MBR_d$; thus with 65 subchannels total, we could instead form 6 groups of 6 homes and give $MBR_d$ to a total of 36 homes.

We plot in Figure~\ref{fig:min_rate_streams_UL}, similar results for the UL for $U \leq N$ with the baseline setting when $\mathcal{R} = 1500$\,m and $T_u = 10$. Once again, we consider the case where the total subchannels in the system is 65, 20, and 10 in Figures~\ref{fig:min_rate_streams_UL_65ch}, \ref{fig:min_rate_streams_UL_20ch} and \ref{fig:min_rate_streams_UL_10ch}, respectively. Recall from \eqref{eq:ul_prb_constraint} that the UL power per PRB $P_{PRB}^k$ is inversely proportional to the number of subchannels allocated to the home. As was the case for the DL, the UL also benefits from multiple streams when the number of homes is small.

\begin{figure}[t]
    \centering
    \includegraphics[width=0.99\linewidth]{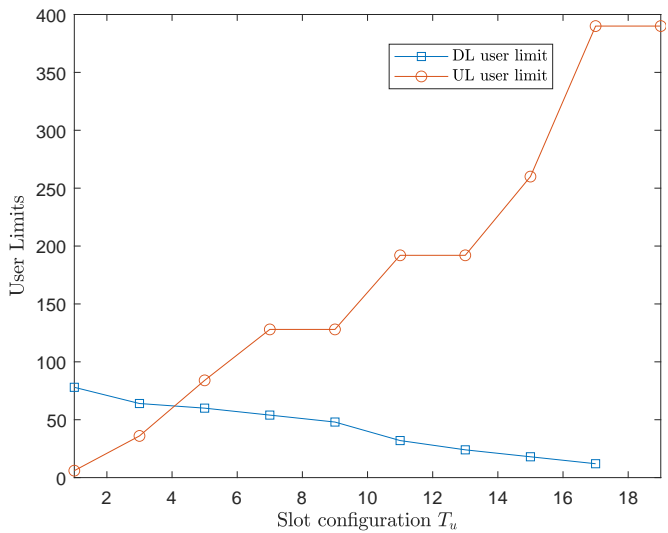}
    \caption{DL and UL user limits for different $T_u$: baseline setting with $\mathcal{R}=1500$\,m.}
    \label{fig:absolute_user_limit_r1500}
\end{figure}

For the UL, we also see that grouping is necessary to serve more than $10$ homes, as we cannot select $U > 10$ homes at once and give them all $MBR_u$ even with 65 subchannels (Figure~\ref{fig:min_rate_streams_UL_65ch}). However, we could easily give 16 homes $MBR_u$ by dividing them into two groups of 8 and giving each group 20 subchannels (Figure~\ref{fig:min_rate_streams_UL_20ch}); or we could do the same with just 10 subchannels available to give each group (Figure~\ref{fig:min_rate_streams_UL_10ch}). Figure~\ref{fig:min_rate_streams_UL_10ch} shows that taking groups of $S_u = 8$ homes and giving each group 10 subchannels would enable 48 homes to receive $MBR_u$ if the total number of channels is 65. Unfortunately, for $T_u = 10$, we cannot offer $MBR_d$ to this many homes.
Hence, for the baseline setting and  $T_u = 10$, we see that the DL is the \emph{bottleneck} of the system. Clearly, decreasing $T_u$ would provide more resources to the DL, which would increase the number of homes that can receive $MBR_d$ while the number of homes that can receive $MBR_u$ would decrease. At a certain point, i.e., a certain value of $T_u$, the UL would become the bottleneck.

\textit{Lessons learned:} From the results in this section we learned that the appropriate number of streams to use depends on the group size. In general more streams are better when the group size is small relative to $N$; however, solving the PD problem enables us to automatically determine the right number of streams. We also showed why user grouping is critical. For group sizes close to $N$, we cannot give every home the MBRs, even if we allocate every subchannel in the system to the group. Thus, to serve $U>N$ homes, it is essential to divide homes into smaller groups and allocate a fraction of the subchannels to each group. These results hold for both the DL and the UL.

\begin{figure}[t]
    \centering
    \includegraphics[width=0.98\linewidth]{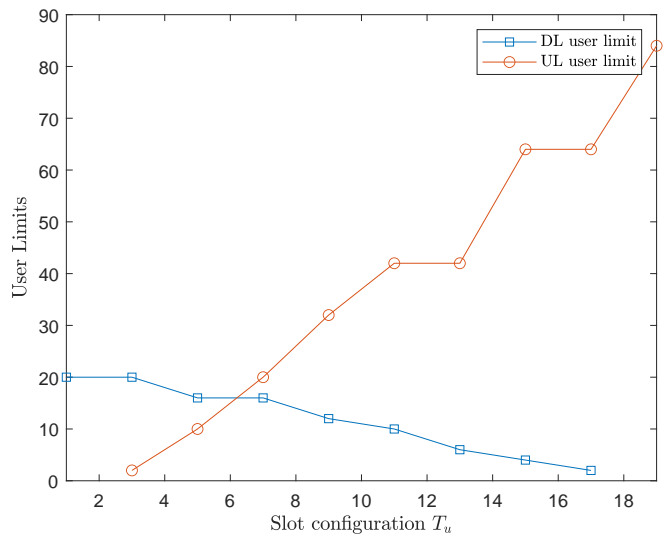}
    \caption{DL and UL user limits for different $T_u$: baseline setting with $\mathcal{R}=5000$\,m.}
    \label{fig:absolute_user_limit_r5000}
\end{figure}

\subsection{Planning Results for Baseline Setting}
Now our goal is to determine, for a given system setting $\mathcal{S}$, the  user limit $U^*(\mathcal{S})$ as well as a configuration $\mathcal{V}^*(\mathcal{S})$ that achieves that limit for a given radius. For this study we set $L = M_{\rm H}$ and allow PD to perform stream selection. We initially study the baseline system  for $|\Omega |= 100$ realizations. Using Algorithm~\ref{alg:ComputeLimit}, we compute the number of subchannels necessary to offer $MBR_d$ and $MBR_u$ for 95\% of realizations for different group sizes and for $T_u \in [1,19]$. Based on these results, we determine the DL user limit $U_d$ and UL user limit $U_u$ for the baseline setting and $\mathcal{R} = 1500$\,m, which we plot in Figure~\ref{fig:absolute_user_limit_r1500} as a function of $T_u$. Per \eqref{eq:AbsLimit}, $U^*$ occurs at the point where the DL and UL user limits intersect. From the figure, we can see that the baseline setting is able to offer the MBR on the DL and UL to 60 homes (i.e., \mbox{$U^* = 60$}) when $\mathcal{R} = 1500$\,m. This happens when  \mbox{$T_u^* = 4$}, i.e., only 20\% of the time-slots are given to the UL. This result is achieved with the optimal group sizes \mbox{$S_d^* = 6$} and \mbox{$S_u^* = 4$} which are both much lower than $N=16$.  

Now we conduct the same study  for $\mathcal{R} = 5000$\,m  (the other setting parameters are as per the baseline) and plot the results in Figure~\ref{fig:absolute_user_limit_r5000}. We obtain \mbox{$U^* = 15$}, \mbox{$T_u^* = 6$}, which are achieved by \mbox{$S_d^* = 4$}, \mbox{$S_u^* = 2$}. Not surprisingly, the user limit is much lower (15 homes) than the one (60 homes) for the original radius of 1500~m and the optimal group sizes are quite small.

\textit{Lessons learned:}
In Table~\ref{tab:limitsBaseline}, we summarize the  user limits and the optimal configurations achieving these limits for the baseline  setting for different cell radii. It is clear that the optimal time-slot and group configuration is dependent on the cell radius. The advantage of user grouping is also clear: for $\mathcal{R} < 5000$~m, user grouping always enables us to give more than $N$ users the MBRs. Furthermore, for all radii except $\mathcal{R}=10000$~m, the downlink is the bottleneck of the system and requires more resources than the uplink (i.e., $T_u^* < T/2$). We observe also that for $\mathcal{R}=10000$~m the problem is infeasible for $S_u \geq 2$; hence, we must resort to SU-MIMO, i.e., $S_u = 1$.

\begin{table}[t]
    \centering
    \begin{tabular}{|c|c|c|c|c|}
        \hline
        $\mathcal{R}$ & $U^*$ & $T^*_u$ & $S^*_d$ & $S^*_u$ \\
         \hline
        1500 & 60 & 4 & 6 & 4 \\
        2000 & 42 & 5 & 6 & 4 \\
        3000 & 32 & 5 & 4 & 4 \\
        4000 & 22 & 6 & 4 & 2 \\
        5000 & 15 & 6 & 4 & 2 \\
        10000 & 3 & 11 & 2 & 1 \\
        \hline
    \end{tabular}
    \caption{User limits for the baseline setting at different radii, with corresponding configuration attaining that limit.}
    \label{tab:limitsBaseline}
\end{table}

\subsection{Validation of Our Approach}\label{sect:validation}
\begin{figure}[t]
    \centering
    \includegraphics[width=\linewidth]{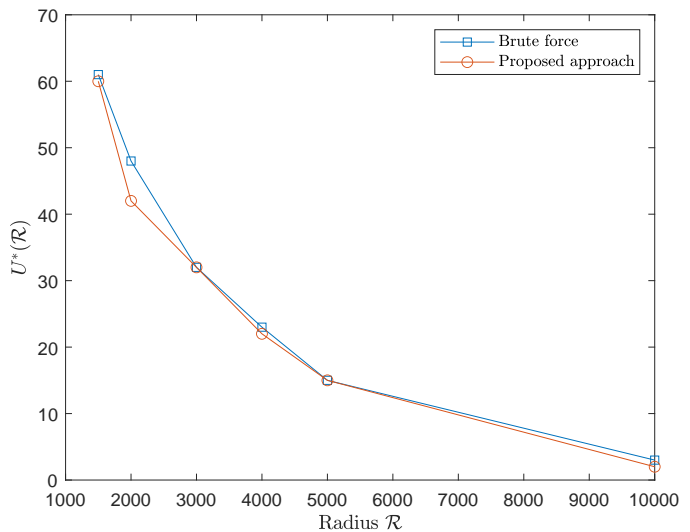}
    \caption{User limits obtained through brute force vs through our proposed approach. Baseline setting, $\Omega = 100$.}
    \label{fig:userLimitBruteForceVsProposedApproach}
\end{figure}

Our simplified approach to determining the user limits gives us \textit{feasible} results (because it is based on solving the approximate PD problems), but we would like to validate our approach to see if the user limits we computed are \textit{reasonable}. To do so, we also compute the user limits by brute force for our baseline setting. We start by generating 100 realizations for $U^*$ homes, where $U^*$ is the user limit obtained by our simple approach, and check if we can indeed give all of the homes the MBRs for 95\% of the realizations (this is always the case). We then start our brute force search for  $X\geq U^*$ homes, generate 100 realizations with $X$ homes and we keep increasing $X$ till we cannot offer the MBRs  to all homes in at least 95 of the 100 realizations. The result of the brute force validation, shown in Figure~\ref{fig:userLimitBruteForceVsProposedApproach}, reveals that in general the ``true'' user limit (computed through brute force) is very close to the one computed using our simple method. In most cases the difference is negligible, however at $\mathcal{R} = 2000$ the difference is about 15\%, which could be explained by some bad luck with the rounding effects.

\textit{Lessons learned:} The beauty of our simple approach is that it gives us a  great feasible solution.

\begin{table*}[t]
    \centering
    \begin{tabular}{|c|ccccccc|}
        \hline
        \multirow{2}{*}{Setting} & \multicolumn{7}{c|}{Parameters $\mathcal{S}$} \\
        \cline{2-8}
         & $M_{\rm BS}$ & $M_{\rm H}$ & $B$ (MHz) & $C$ & $P_{\rm max}$ (W) & $MBR_d$ (Mbps) & $MBR_u$ (Mbps) \\
        \hline
        Baseline            & 64  & 4 & 25 & 65 & 40 & 30 & 5 \\
        Double bandwidth    & 64  & 4 & \textbf{50} & \textbf{133} & \textbf{80} & 30 & 5 \\ 
        Double BS antennas  & \textbf{128} & 4 & 25 & 65 & 40 & 30 & 5 \\
        Double home antennas & 64 & \textbf{8} & 25 & 65 & 40 & 30 & 5 \\
        Double BS power     & 64 & 4 & 25 & 65 & \textbf{80} & 30 & 5 \\
        Half MBR            & 64 & 4 & 25 & 65 & 40 & \textbf{15} & \textbf{2.5} \\
        \hline
    \end{tabular}
    \caption{System settings we consider. Parameters that differ from the baseline are in bold.}
    \label{tab:settings}
\end{table*}

\subsection{Network Planning Results}

Now we examine the impact of several other systems settings, which are summarized in Table~\ref{tab:settings}. These settings are considered relative to the baseline: double the overall bandwidth $B$, double the antennas at the BS $M_{\rm BS}$, double the antennas at the homes $M_{\rm H}$, double the power at the BS $P_{\rm max}$, and half $MBR_d$ and $MBR_u$. As we did for the baseline, we will study the user limits and optimal configurations achieving those limits for different cell radii, i.e.,  $\mathcal{R} \in \{1500,2000,3000,4000,5000,10000\}$\, (meters).

To reduce the overall computations, we make the following observations a priori: 
\begin{enumerate}
    \item For the double BS power setting, we may reuse the UL results from the baseline scenario because changing $P_{\rm max}$ has no impact on the UL user limit (problem $\mathbf{P_u}$ is not dependent on $P_{\rm max}$).
    \item For the double bandwidth setting, for a given $\mathcal{V}_u=(T_u,S_u)$ we may reuse $n(\mathcal{V}_u,L)$ (the subchannels necessary to achieve $MBR_u$) from the baseline (from constraint \eqref{eq:ul_prb_constraint}, the results of $\mathbf{P_u}$ are determined by $|\mathcal{C}_u(\hat{\mathcal{U}})|$ and not by the total channels $C$); then we may use \eqref{eq:ulLimit} to recompute $U_u(\mathcal{V}_u,L,C)$ with the larger value of $C$.
\end{enumerate}

\begin{figure*}[t]
    \centering
        \begin{minipage}{0.33\linewidth}
        \centering
        \includegraphics[width=0.98\textwidth]{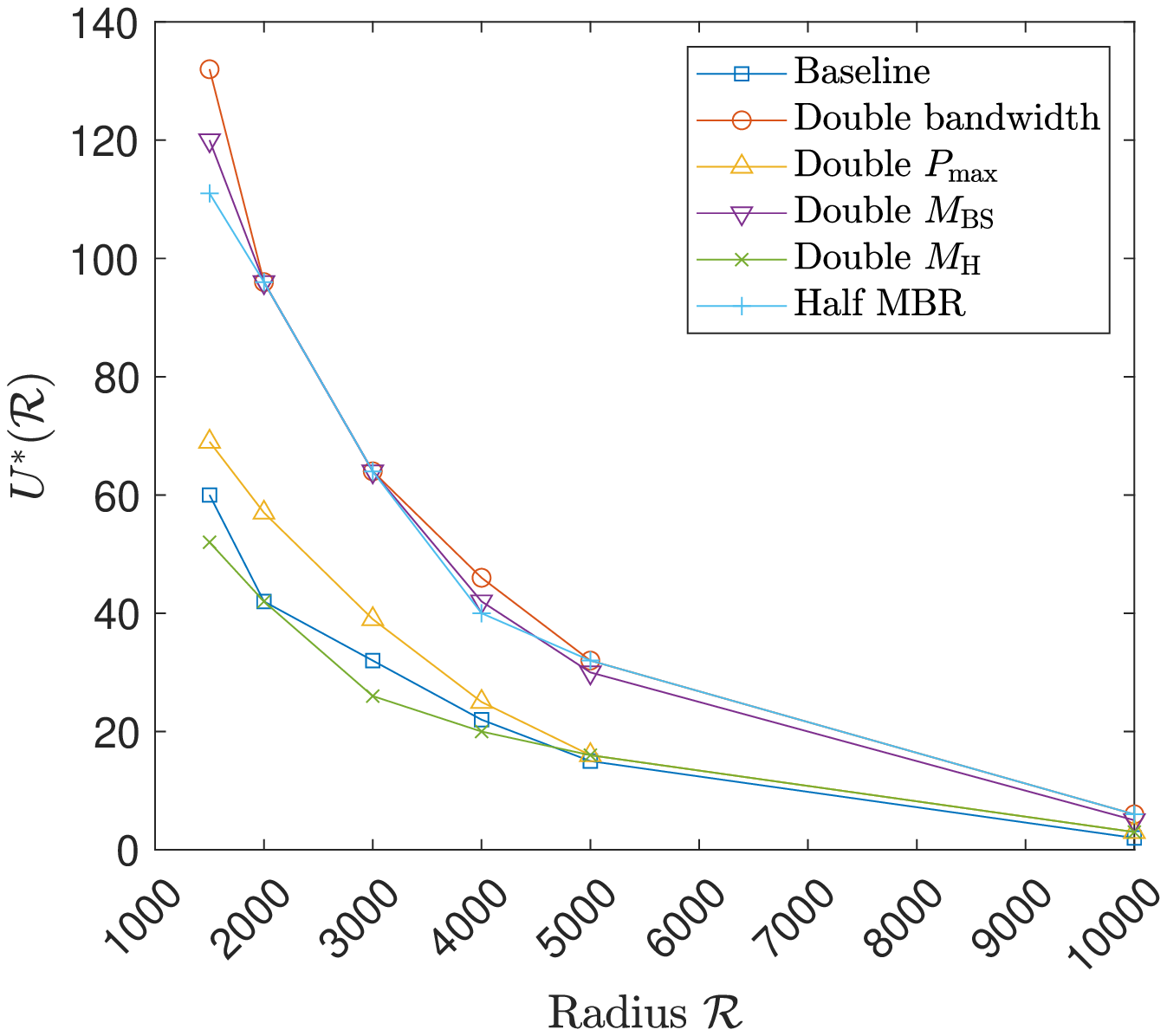}
        \subcaption{Absolute user limits.\\~}\label{fig:absoluteUserLimits}
    \end{minipage}
    \begin{minipage}{0.32\linewidth}
        \centering
        \includegraphics[width=0.98\textwidth]{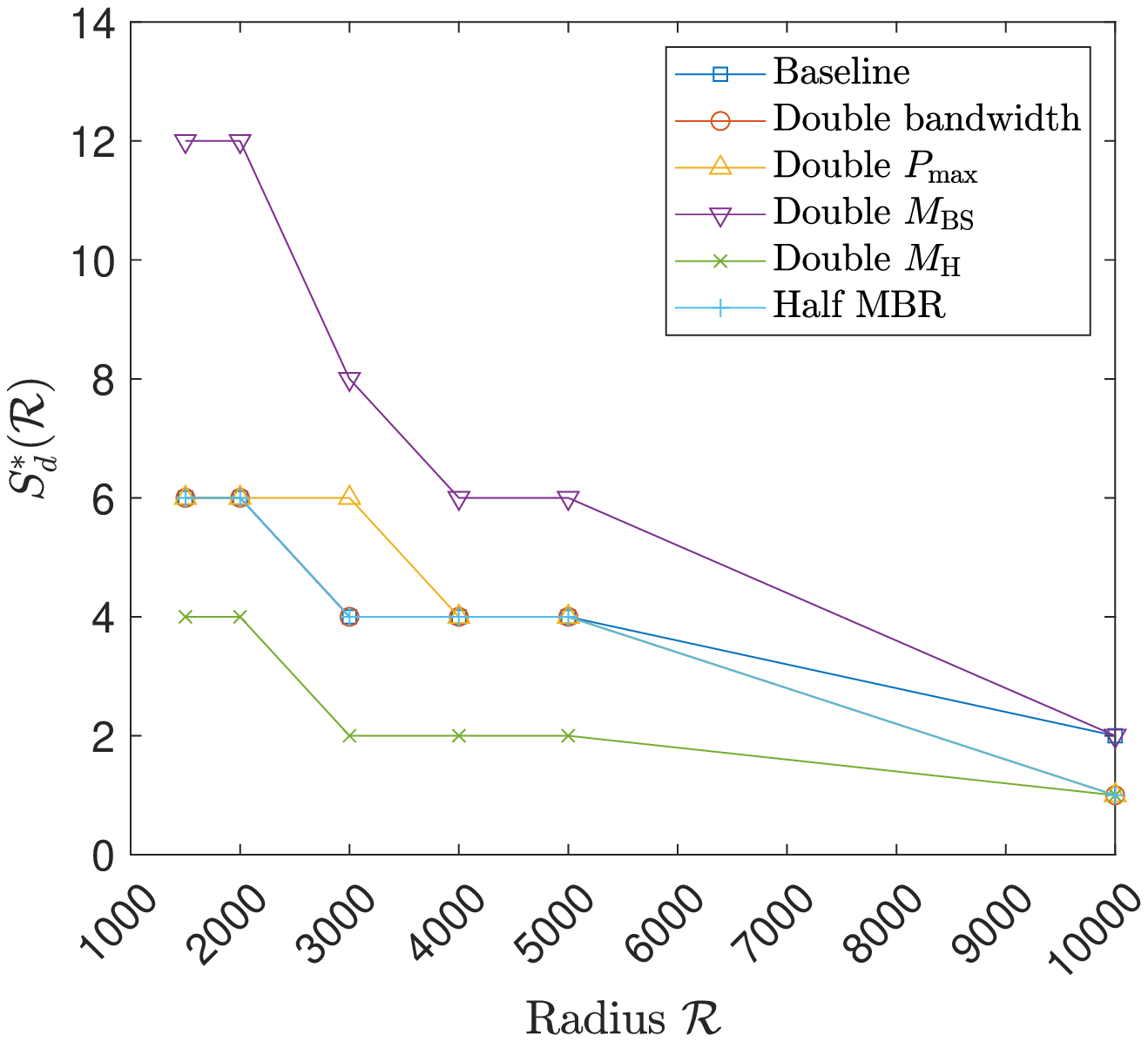}
        \subcaption{Optimal DL group sizes. Baseline, double bandwidth, half MBR overlap for $\mathcal{R} \leq 5000$.
        }\label{fig:idealDLGroups}
    \end{minipage}
    \begin{minipage}{0.32\linewidth}
        \centering
        \includegraphics[width=0.98\textwidth]{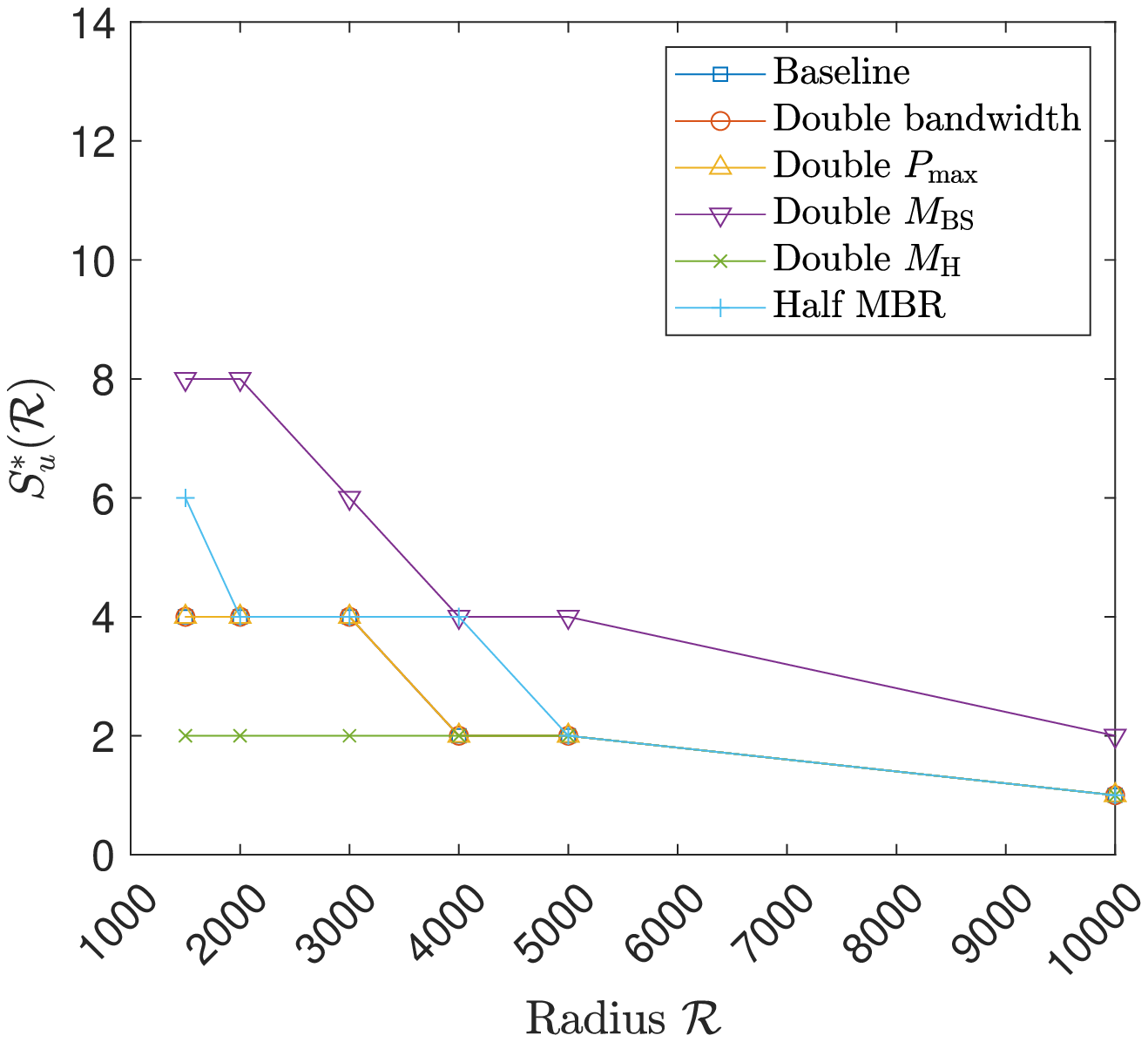}
        \subcaption{Optimal UL group sizes. Baseline, double bandwidth, double power overlap for all $\mathcal{R}$.}\label{fig:idealULGroups}
    \end{minipage}
    \caption{Absolute user limits and group sizes required to achieve the limits for different settings.}
    \label{fig:preplanningLimits}
\end{figure*}

We plot the  user limits for each setting in Figure~\ref{fig:absoluteUserLimits} as a function of the radius. We observe that the user limit can be doubled when the bandwidth $B$ is doubled, when the number of BS antennas $M_{\rm BS}$ is doubled, or when $MBR_d$ and $MBR_u$ are halved. Adding antennas at the BS might be a solution when an operator is limited in the bandwidth it has at its disposal. On the other hand, doubling the BS power $P_{\rm max}$ alone only provides a marginal increase to the total number of homes that can be given the MBRs. Counter intuitively, increasing the antennas at the homes adversely affects the number of homes that can receive the MBR for small cell radii and does not improve it for large radii (the reason for this is explained below).

In Figure~\ref{fig:idealDLGroups} and Figure~\ref{fig:idealULGroups}, we plot the group sizes achieving the user limits in the DL and UL, respectively. Note that, in the DL, the baseline, double bandwidth, and half MBR settings achieve their user limits with the same DL group sizes at all radii except $\mathcal{R}=100000$; in the UL, the baseline, double bandwidth, and double power settings achieve their limits with the same UL group sizes at all radii. When $M_{\rm BS} = 128$ and $M_{\rm H} = 4$ we have $N = 32$ and, hence, we can form larger groups. Indeed, Figures~\ref{fig:idealDLGroups} and \ref{fig:idealULGroups} show that this setting achieves the DL and UL user limits with larger group sizes than the other settings.

\begin{figure}[t]
    \centering
    \includegraphics[width=\linewidth]{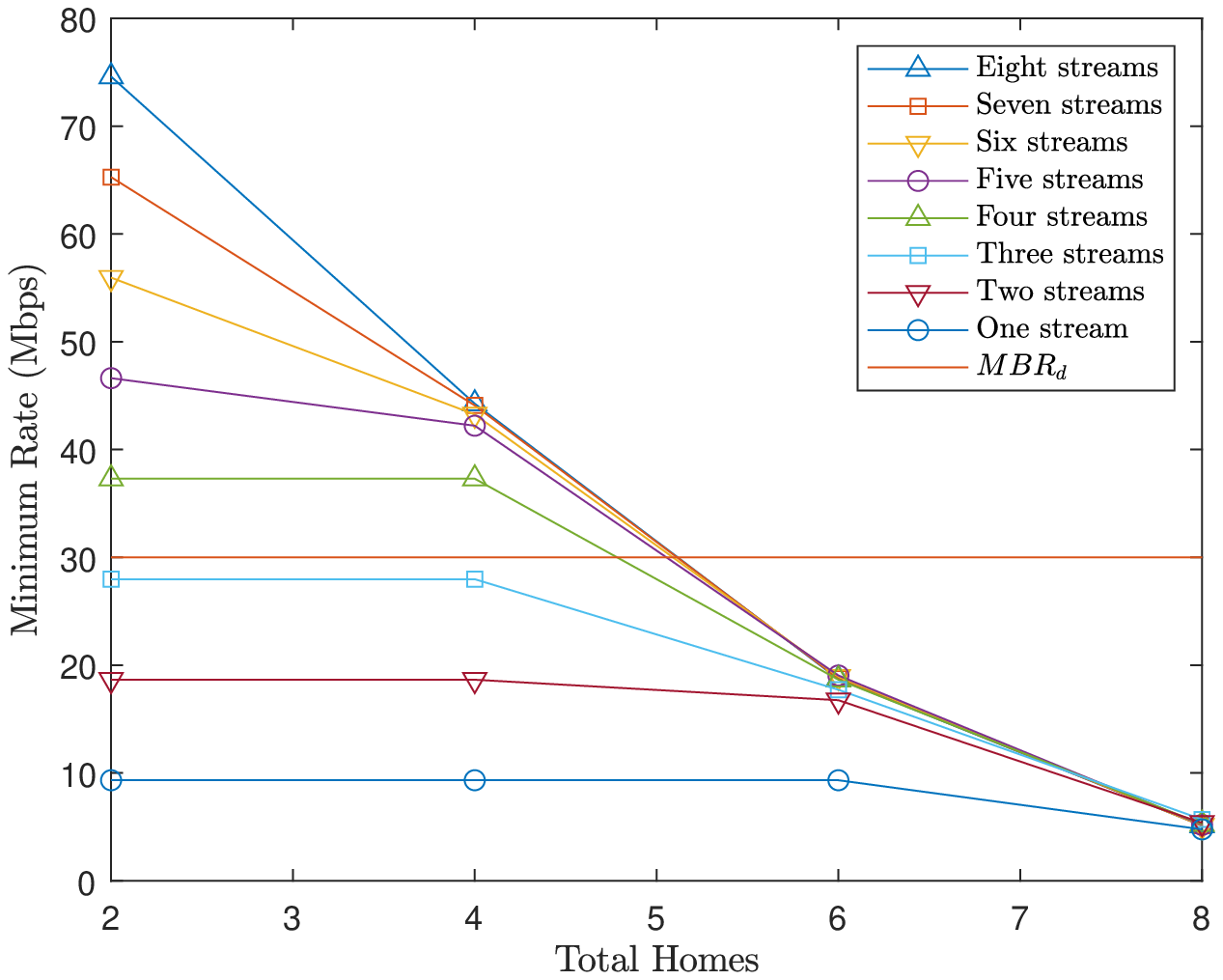}
    \caption{Minimum DL user rates for $U \leq N$: \mbox{$\Omega = 100$}, \mbox{$\mathcal{R} = 1500$\,m}, \mbox{$M_{\rm BS} = 64$}, \mbox{$M_{\rm H} = 8$}, \mbox{$P_{\rm max} = 40$\,W}, \mbox{$MBR_d = 30$\,Mbps, $T_u = 10$} (10 subchannels are allocated to the homes, but $P_{PRB}^d = \frac{P_{\rm max}}{65}$).}
    \label{fig:min_rate_streams_DL_MH8}
\end{figure}

In the case where the number of antennas at the home are doubled, we have $M_{\rm H} = 8$. With $M_{\rm BS} = 64$ this gives $N = 8$, meaning fewer homes can be selected together. Even though $M_{\rm H} = 8$ increases the maximum number of streams per home to 8, in order to take advantage of these extra streams we are forced to use smaller group sizes, as is shown in Figure~\ref{fig:min_rate_streams_DL_MH8} for the DL when $T_u = 10$ and 10 suchannels are allocated per group. Despite the fact that we can gain in througphut at these smaller group sizes compared to when $M_{\rm H} = 4$ (Figure~\ref{fig:min_rate_streams_DL_10ch}), we lose out in the channels that are available to distribute among enough groups to boost the user limit. Thus, for the same number of subchannels that we could give a group of 6 homes the MBRs when $M_{\rm H} = 4$, we can only give the MBRs to a smaller group of homes when $M_{\rm H} = 8$; hence, under this setting, the user limit is less than the baseline and $S^*_d$ and $S^*_u$ are typically smaller as shown in Figures~\ref{fig:idealDLGroups} and \ref{fig:idealULGroups}, respectively. Note that when $\mathcal{R}=10000$~m, for all settings except the one where $M_{\rm BS} = 128$, the problem is only feasible if we treat the UL as SU-MIMO (i.e., with $S_u = 1$).

\textit{Lessons learned:} The attainable user limits are highly dependent on the system settings. As one might expect, doubling the system bandwidth or halving the MBRs means that twice as many homes can be supported; however, operators with limited spectrum resources could also achieve similar results by doubling the antennas at the BS instead. On the other hand, increasing the BS transmit power provides minimal improvements, especially for large  cell radius. We also see that there is a close relationship between the optimal group sizes and the number of antennas at the BS and at the homes. Increasing $M_{\rm BS}$ increases $N$ and enables larger group sizes to be used, supporting more users in the system. Increasing $M_{\rm H}$ increases the maximum possible number of streams; however, it also decreases $N$, thus limiting the size of groups that can be used and counteracting the effects of more streams.

\subsection{A Word on Runtime and Complexity}
In order to compute the user limit, we need to solve multiple instances of the PD problem on the DL and on the UL. Indeed, for a given realization, we try different values of $T_u$, different values of the group sizes $S_d$ and $S_u$, and of course we redo this for 100 realizations. Hence, we need to be able to solve the PD problem quite fast. Solving one instance of the original PD problem $\mathbf{P_d^0}$ is very long compared to the approximated problem $\mathbf{P_d}$. This is because $\mathbf{P_d^0}$ uses the practical MCS function, introducing $|\mathcal{C}_d(\hat{\mathcal{U}})| |\mathcal{T}_d||\hat{\mathcal{U}}|LQ$ additional (integer) variables $\{I_{k,l,q}^{c,t}\}$ (one variable per MCS level for each stream of every user in each PRB) to be solved compared to problem $\mathbf{P_d}$. The practical MCS function is piecewise constant, which makes $\mathbf{P_d^0}$ a non-convex MINLP. This type of problem is difficult to solve. It can be solved by advanced commercial solvers (e.g. BARON) that employ branch-and-bound methods, but these techniques have exponential complexity in general~\cite{zhang1996branch}. Practically, when we used BARON for $\mathbf{P^0_d}$, we were restricted in the size of the problem we could solve. By replacing the practical MCS function with a continuous concave approximation in $\mathbf{P_d}$, we make the PD problem into a convex NLP, which can be solved through more efficient techniques, e.g. interior-point methods~\cite{boyd2004convex}. $\mathbf{P_d}$ also has much less variables. Practically, for a given mid-size system, $\mathbf{P_d}$ can be solved in minutes while $\mathbf{P^0_d}$ takes hours to be solved if at all. In this paper, we use a tight, continuous approximation for the MCS function to get very good feasible solutions for the original problem.\footnote{In this approximation, the values of $a$ and $b$ in \eqref{eq:ApproxMCS} are selected to minimize the mean squared error between the practical and continuous MCS functions. These functions are illustrated in Fig.~1 of \cite{zhang2021joint}.} If, instead, we take an approximation that upper bounds the MCS function, we would get an upper bound on the objective function. We have checked that the feasible solution that we obtain with the tight approximation yields an objective function that is less than 5\% from the upper bound, thus validating our approach~\cite{zhang2021joint}.

\begin{figure}[t]
    \centering
    \includegraphics[width=0.48\textwidth]{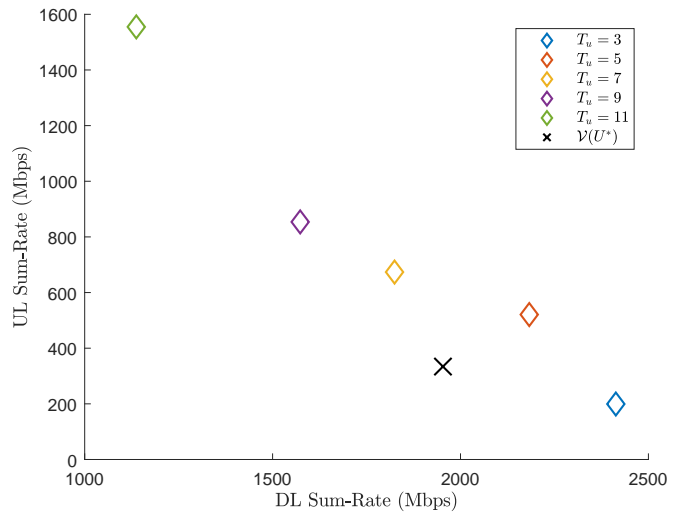}
    \caption{Achievable sum-rates for $U=30$ homes for the baseline setting when $\mathcal{R}=1500$~m. The rate achieved by the planning configuration $\mathcal{V}(U^*)$ (where $U^* = 60$) is shown with a $\times$.}
    \label{fig:feasibleRates1500}
\end{figure}

\subsection{Insights for Network Operation}\label{sect:operation}
The network planning results provide the user limits (i.e., the maximum number of homes that can be given the MBRs) for different system settings $\mathcal{S}$ as well as the configurations $\mathcal{V}$ necessary to attain those limits. To obtain these results we assumed every home was always active; however, typically, even if an MNO has $U^*$ homes subscribing to its FWA service in a given cell, the number of active homes $U$ would  be less than the limit $U^*$ most of the time. The results obtained through the network planning study above can yield information about how the network should be operated  when $U < U^*$, i.e., should the configuration remain the same and equal to what we call in the following the ``planning configuration'' $\mathcal{V}(U^*)=(T_u(U^*),S_d(U^*),S_u(U^*))$ at all times or should it change depending on $U$?

This question is linked to the following one: if many configurations are $U$-feasible, which one is the best? This, of course, depends on an additional criteria defining ``best". We decided to use the  sum-rate (averaged over the 95\% realizations that we keep) as that criteria. For example, in Figure~\ref{fig:feasibleRates1500}, we show the DL and UL sum-rates that can be attained when $U = 30$ homes are active under the baseline setting and $\mathcal{R} = 1500$\,m. Each point in the plot corresponds to the $U$-feasible configuration providing the largest sum-rates on the DL and the UL for a given $T_u$. From this data it is clear that different configurations can provide different performance on the DL and UL. Moreover, for $U =30$ there exist $U$-feasible configurations that can provide better sum-rates than $\mathcal{V}(U^*)$ both on the DL \emph{and} the UL. To determine which configuration is best, we need to decide how much weight the MNO gives to the DL and UL sum-rates. Let $\alpha$ be the DL weight, and define the weighted sum-rate for $U\leq U^*$ and configuration $\mathcal{V}$

\begin{align}
    \begin{multlined}
    WSR(\alpha,U,\mathcal{V}) = \alpha \times SR_d(U,\mathcal{V}_d) + \\ (1-\alpha) \times SR_u(U,\mathcal{V}_u) \label{eq:sumrate}
    \end{multlined}
\end{align}
For $U=30$, the baseline setting and $\mathcal{R}=1500$~m, if $\alpha=0.5$ (resp. $\alpha=0.75$), the best WSR is 1352~Mbps (resp. 1860~Mbps) which is attained with $T_u = 5$ (resp. $T_u = 3$) while if we had used the configuration obtained via planning (i.e., to enable $U^*=60$), the WSR would have been 1144~Mbps (resp. 1548~Mbps). We found that using a configuration for a number of active homes $U\leq U^*$ could yield performance of up to 30\% better than with the planning configuration depending on the values of $U$, $\alpha$ and the radius among other things.

\begin{figure}[t]
    \centering
    \includegraphics[width=0.47\textwidth]{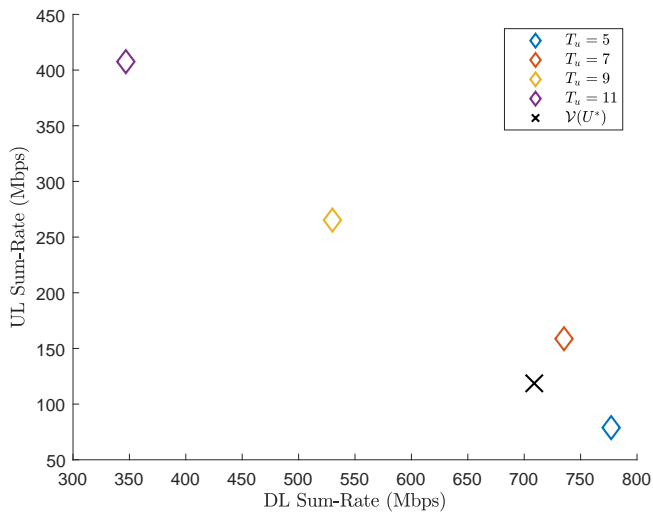}
    \caption{Achievable sum-rates for $U=8$ homes for the baseline setting when $\mathcal{R}=5000$~m. The rate achieved by the planning configuration $\mathcal{V}(U^*)$ (where $U^* = 15$) is shown with a $\times$.}
    \label{fig:feasibleRates5000}
\end{figure}

We repeat the analysis for the baseline setting for $\mathcal{R} = 5000$\,m and $U = 8$ ($U^*=15$ for this setting and radius) and 
plot the sum-rates that can be achieved for different $T_u$ in Figure~\ref{fig:feasibleRates5000}. In this case there are fewer configurations that are $U$-feasible, and less configurations that can improve upon $\mathcal{V}(U^*)$ both for the UL and the DL. Indeed, for $\alpha = 0.5$ (resp. $\alpha = 0.75$) the best WSR is 447~Mbps (resp. 603~Mbps), which is attained with $T_u = 7$ (resp. $T_u = 5$), whereas $\mathcal{V}(U^*)$ yields a WSR of 413~Mbps (resp. 561~Mbps). At this radius the performance improvement due to choosing a better configuration for $U\leq U^*$ is limited to 8\%.

\textit{Lessons learned}: From these results, we learn that by adapting the configuration to the number of active homes,  we might provide noticeable performance improvements to homes, especially if the radius is not too large. In the past, dynamic resource sharing in TDD systems was not practical, and TDD-LTE systems were typically statically or semi-statically configured. However 5G provides significantly more flexibility in this regard, making it possible to change time-slot configurations frame-to-frame \cite{3gppTS38213}. This flexibility is designed to make it possible for operators to serve diverse, sometimes conflicting user requirements, e.g. DL-heavy eMBB vs UL-heavy mMTC. Our results show that such a scheme could also be used to dynamically operate the network based on the number of active homes. Dynamic TDD could create interference challenges between neighbouring cells and there is ongoing research in this area \cite{gupta2016rate,guo2018dynamic}, however this interference should be easier to manage in rural environments.

\section{Conclusion}\label{sect:conclusion}

In this paper we have studied the problem of planning a rural 5G-TDD MU-MIMO network to jointly provide broadband services (and minimum bit rates) in the DL and UL to FWA homes under block diagonalization (BD) precoding and combining. We first studied the case where the total number of homes $U \leq N = \left\lceil \frac{M_{\rm BS}}{M_{\rm H}} \right\rceil$ (the ratio of antennas at the BS to the antennas at one home) and showed that multi-antenna FWA homes benefit from using multiple data streams. Moreover, we showed that dividing homes into static groups with fixed subchannel allocations  ensures many more homes can be given the MBR than selecting close to $N$ homes at the same time.

Next we studied the case where $U > N$. Since the system is TDD, determining the maximum number of homes that can be given MBRs jointly in the DL and the UL would require solving the DL and UL problems many times for different total numbers of homes $U$, different group sizes, and different slot configurations. To simplify this process, we developed a simple procedure to determine the DL and UL user limits for a given time-slot configuration and a given group size, which we validated against brute force computation. Under this procedure we analyzed the impact of user grouping on the DL and UL user limits for different time-slot configurations and determined the absolute user limits for a number of different system settings at different cell radii. Finally, we provided some insights on how networks can be dynamically operated to provide DL and UL MBRs to FWA homes. 

The tools that we have developed and the results that we have obtained can help MNOs decide if introducing FWA for fixed broadband is profitable, i.e., if the user limits are large enough to warrant the implementation of that service. Furthermore, the results can help MNOs provision their networks to meet a certain user limit.

\ifCLASSOPTIONcaptionsoff
  \newpage
\fi

\bibliographystyle{IEEEtran}
\bibliography{IEEEabrv,bibtex/bib/main}

\end{document}